\documentclass[letterpaper,12pt]{article}

\usepackage{amssymb,latexsym,amsmath}

\usepackage{fullpage}
\usepackage{nicefrac}
\usepackage{mathrsfs}

\usepackage{wrapfig}

\usepackage[pdftex]{graphicx}

\makeatletter \@addtoreset{equation}{section}
\makeatother

\begin{document}

\begin{titlepage}
	\thispagestyle{empty}
	\begin{flushright}
		\hfill{DFPD-11/TH/07}
	\end{flushright}
	
	\vspace{35pt}
	
	\begin{center}
	    { \LARGE{\bf Black holes in supergravity: \\[2mm] flow equations and duality}}
		
		\vspace{50pt}
		
		{Gianguido Dall'Agata}
		
		\vspace{25pt}
		
		{
		{\it  Dipartimento di Fisica ``Galileo Galilei''\\
		Universit\`a di Padova, Via Marzolo 8, 35131 Padova, Italy}
		
		\vspace{15pt}
		
		\emph{and}
		
		\vspace{15pt}
		
		{\it   INFN, Sezione di Padova \\
		Via Marzolo 8, 35131 Padova, Italy}}		
		
		\vspace{60pt}
		
		\emph{Based on lectures given at the School on Attractor Mechanism 2009 (LNF, Italy), at the 27th Nordic Spring String Meeting (NBI Copenhagen, Denmark) and at the Black Objects in Supergravity School 2011 (LNF, Italy)}
	
		\vspace{40pt}
		
		{ABSTRACT}
	\end{center}
	
	\vspace{10pt}
	
We review the physics of extremal black holes in supergravity theories, emphasizing the role of the first order formalism underlying single-centre solutions, the attractor mechanism and describing the recent progress in constructing general non-supersymmetric multi-centre configurations.

\end{titlepage}

\baselineskip 6 mm



\section{Introduction} 
\label{sec:introduction}

The analysis of black hole solutions and the study of their physics is an active and important branch of contemporary theoretical physics.
In fact, not only black holes are an excellent theoretical laboratory for understanding some features of quantum gravity, but they can also be successfully used as a tool in applications to nuclear physics, condensed matter, algebraic geometry and  atomic physics.
For this reason, black holes are considered the ``Hydrogen atom'' of quantum gravity \cite{Maldacena:1996ky} or the ``harmonic oscillator of the 21st century'' \cite{Stromingertalk}.

The existence of black holes seems to be an unavoidable consequence of General Relativity (GR) and of its extensions (like supergravity). 
Classically, the horizon of black holes protects the physics in the outer region from what happens in the vicinity of singular field configurations that can arise in GR from smooth initial data.
However, already at the semiclassical level, black holes emit particles with a thermal spectrum \cite{Bekenstein:1973ur,Hawking:1971tu}.
A thermodynamic behaviour can also be associated to black holes from the laws governing their mechanics \cite{Townsend:1997ku} and, in particular, one can associate to a black hole an entropy $S$ proportional to the area $A$ of its event horizon (measured in Planck units $l_P^2 = G\hbar/c^3$)
\begin{equation}
	S = \frac{k_B}{l_P^2}\;\frac{A}{4}.
\end{equation}
In most physical systems the thermodynamic entropy has a statistical interpretation in terms of counting microscopic configurations with the same macroscopic properties, and in most cases this counting requires an understanding of the quantum degrees of freedom of the system. 
The identification of the degrees of freedom that the Bekenstein--Hawking entropy is counting is a long-standing puzzle that motivated much theoretical work of the last few years.
String Theory, being a theory of quantum gravity, should be able to provide a microscopic description of black holes and hence justify Bekenstein--Hawking's formula.
By now we have strong indications and many different and compelling examples where String Theory successfully accomplishes this goal, although often simplifying assumptions are made so that the configurations which are considered are not very realistic.
In particular, black holes are non-perturbative objects and only for special classes of solutions (mainly supersymmetric) string theory at weak coupling can reproduce the correct answer\footnote{Recently there has been also a lot of progress in understanding the nature of the entropy for Kerr black holes and close to extremal examples of this sort can be realized in nature \cite{Bredberg:2011hp}.} \cite{Strominger:1996sh,David:2002wn,Pioline:2006ni}.
However, there is now a growing evidence that also for non-zero coupling we can identify candidate microstate geometries, whose quantization may eventually yield an entropy that has the same parametric dependence on the charges as that of supersymmetric black holes \cite{Mathur:2005zp,Kanitscheider:2006zf,Bena:2007kg,Balasubramanian:2008da}.

In the last few years a lot of progress has been made in understanding the physics of \emph{extremal non-supersymmetric solutions} and of their candidate microstates.
The aim of these lectures is to provide an elementary and self-contained introduction to supergravity black holes, describing in detail the techniques that allow to construct full extremal solutions and to discuss their physical properties.
We will especially focus on the peculiar role of scalar fields in supergravity models and on the flow equations driving them to the attractor point provided by the black hole horizon.
We will also discuss the multicentre solutions and the role of duality transformations in establishing the classes of independent solutions.


\section{Black holes and extremality} 
\label{sec:extremality}

In this section we will review some general properties of black holes and discuss the concept of \emph{extremality}, both in the context of geometrical and of thermodynamical properties of the solutions.

\bigskip

We will be interested in charged black hole configurations, so our starting point is the Einstein--Maxwell action in 4 dimensions, with Lagrangian density given by
\begin{equation}
 e^{-1} {\cal L} = R - \frac14\, F_{\mu\nu}F^{\mu\nu}.
\label{basiclagrangian}
\end{equation}
For the sake of simplicity we will look for static, spherically symmetric and charged solutions.
This means that the line element describing the metric should be of the form
\begin{equation}
  ds^2 = - e^{2U(r)}dt^2 + e^{-2U(r)}dr^2 +  r^2 d \Omega^2,
\end{equation}
where $d \Omega^2 = d \theta^2 + \sin^2 \theta\, d \phi^2$ is the line element of a two-sphere and $U$ is the warp factor, which depends only on the radial variable in order to respect spherical symmetry.
For the same reason, the 2-form associated to the Maxwell field $F_{\mu\nu}$ should be of the form
\begin{equation}
  F = P \, \sin \theta\, d \theta \wedge d \phi + Q \,dt \wedge \frac{dr}{r^2},
\end{equation}
so that, by integrating over a sphere, one gets the electric and magnetic charge of the configuration:
\begin{equation}
  \frac{1}{4\pi} \int_{S^2} F = P\,, \qquad   \qquad \frac{1}{4\pi} \int_{S^2} \star F = Q.
\end{equation}
By solving the equations of motion derived from (\ref{basiclagrangian}) we obtain the following expression for the warp factor
\begin{equation}
  e^{2U(r)} = 1 - \frac{2 M}{r} + \frac{P^2+Q^2}{r^2},
\end{equation}
which is the appropriate one for a Reissner--Nordstr\"om black hole and reduces to the one by Schwarzschild for $P=Q=0$.

The solution above contains a singularity at $r = 0$, as one can see by computing the quadratic scalar constructed in terms of the Ricci tensor
\begin{equation}
  R_{\mu\nu}R^{\mu\nu} = 4 \frac{(Q^2+P^2)^2}{r^8} \stackrel{r\to 0}{\longrightarrow} \infty
\end{equation}
(For the special case $P=Q=0$ we can still find a singularity in $R_{\mu\nu \rho \sigma}R^{\mu\nu \rho \sigma} = 48 \frac{M^2}{r^6}$).
However, the singularity is hidden by the horizons appearing at the zeros of the warp-factor function
\begin{equation}
  e^{2U} = 0 \quad \Leftrightarrow \quad r_{\pm} = M \pm \sqrt{M^2 - (P^2 + Q^2)}.
\end{equation}
The two solutions are real as long as $M^2 \geq  P^2 + Q^2$, while the singularity becomes naked for smaller values of the mass.
This means that, \emph{for fixed charges, there is a  minimum value of the mass for which the singularity is screened by the horizons}.
At such value the warp factor has a double zero, the two horizons coincide and the semi-positive definite parameter
\begin{equation}
	\label{cdef}
  c = r_+-r_- = \sqrt{M^2 - (P^2 + Q^2)}\,,
\end{equation}
which we introduce for convenience, is vanishing.
The corresponding black hole configuration is called \emph{extremal} ($c=0$ or $M = \sqrt{P^2 + Q^2}$).
Note that in the uncharged limit $c=M$, which is the extremality parameter for the Schwarzschild solution.
This means that extremal Schwarzschild black holes are necessarily \textit{small}, i.e.~with vanishing horizon area at tree level.

Although the singularity is timelike (for charged solutions) and hence one can interpret it as the presence of a source, the existence of the horizons guarantees that the physics outside the horizon is not influenced by what happens inside, where one meets the singularity.
This fact is easily seen by computing the time it takes for a light ray traveling radially to reach the horizon from infinity, as measured by an observer sitting far from the black hole.
By taking $ds = 0$ for constant $\theta$ and $\phi$ one gets that
\begin{equation}
  \sqrt{g_{tt}}\, dt = \sqrt{g_{rr}} \,dr,
\end{equation}
so that the time it takes for a light ray to travel radially between two points at distance $r_1$ and $r_2$ from the singularity is proportional to the distance measured with a weight given by the inverse of the warp factor
\begin{equation}
	t_{12} = \int_{r_1}^{r_2} \sqrt{\frac{g_{rr}}{g_{tt}}} d\tilde r = \int_{r_1}^{r_2} e^{-2U(\tilde r)} d\tilde r.
\end{equation}
This expression goes to infinity when $r_1 \to r_+$ and therefore a signal from the horizon takes an infinite time to reach a far distant observer.

\begin{figure}
	[htb] \centerline{  
	\includegraphics[scale=1.6]{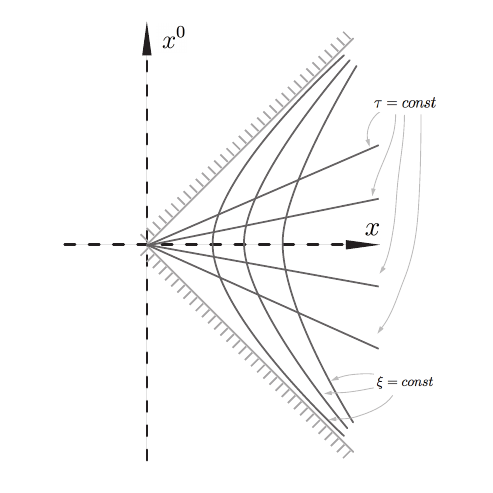} 
	\includegraphics[scale=1.6]{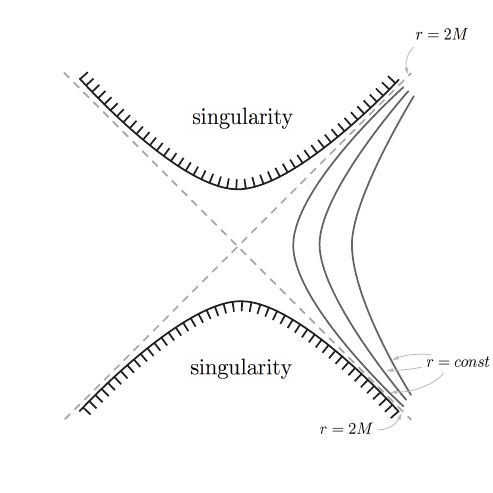} } \caption{Minkowski and Schwarzschild spacetimes in Rindler coordinates. The first diagram approximates the second close to the horizon.} \label{rindlerplots} 
\end{figure}

The physics close to the horizon can be better understood by considering the expansion of the solution obtained above for $r$ close to $r_+$.
The only non-trivial function in the metric is given by the warp factor, which approaches
\begin{equation}
	\label{warpnonextremal}
  e^{2U} = \frac{(r-r_+)(r-r_-)}{r^2} \quad \stackrel{r\to r_+}{\longrightarrow} \quad\frac{r_+-r_-}{r_+^2}\; \rho,
\end{equation}
where we introduced a new coordinate $\rho$ measuring the distance from the outer horizon: $\rho = r- r_+$.
The resulting near horizon geometry is 
\begin{equation}
  ds^2 \to - \frac{r_+ - r_-}{r_+^2} \rho dt^2  + \frac{r_+^2}{r_+-r_-} \frac{d \rho^2}{\rho} + r_+^2\, d \Omega^2,
\end{equation}
which can be interpreted as the product of a 2-dimensional Rindler spacetime with a 2-sphere of radius $r_+$.
We can actually make this result explicit by performing another change of coordinates $(t,\rho)\mapsto (\tau, \xi)$ as follows
\begin{equation}
  \rho = e^{2 \alpha \xi}, \qquad t = \frac{1}{4 \alpha^2}\, \tau, \qquad \alpha = \frac{\sqrt{r_+-r_-}}{2 r_+}.
\end{equation}
This leads to a near-horizon metric described by
\begin{equation}
	\label{NHnonextremal}
  ds^2_{NH} = e^{2 \alpha \xi}\left(- d \tau^2 + d \xi^2\right) + r_+^2 \, d \Omega^2.
\end{equation}
The geometry of the non-compact part is 2-dimensional Minkowski spacetime as seen by an observer that is uniformly accelerated with acceleration $\alpha = \sqrt{\alpha_\mu \alpha^\mu}$.
In fact the change of coordinates from the standard ones to Rindler's is dictated by the trajectory of an accelerated observer
\begin{equation}
	\label{trajectory}
  x(x^0) = \frac1\alpha \sqrt{1 + \alpha^2 (x^0)^2}\,,
\end{equation}
and $\tau$ denotes the proper time 
\begin{equation}
	x^0(\tau) = \frac1\alpha \sinh( \alpha \tau).
\end{equation}
Our derivation explains this acceleration as the effect of gravitation and one can actually show that $\alpha$ coincides with the surface gravity of the black hole.
In fact surface gravity is given in terms of the derivative of the null Killing vector generating the horizon surface, computed at the surface \cite{Townsend:1997ku}
\begin{equation}
	\alpha^2 = \left[-\frac12 \nabla_\mu \xi_\nu \nabla^\mu \xi^\nu\right]_{r = r_+}
\end{equation}
and the two expressions coincide.
 
\subsection{Thermodynamics} 
\label{sub:thermodynamics}

Hawking and Unruh showed that an accelerated observer following the trajectory described in (\ref{trajectory}) sees a thermal spectrum with temperature proportional to the acceleration:
\begin{equation}
  T = \frac{\alpha}{2 \pi}.
\end{equation}
A simple heuristic argument to understand this result follows by considering the near-horizon geometry rotated to Euclidean signature.
This is the same procedure that is used to describe quantum field theories with temperature.
In fact, in quantum mechanics, for a system with Hamiltonian $H$, the thermal partition function is 
\begin{equation}
	Z = {\rm Tr} \, e^{-\beta H},
\end{equation}
where $\beta$ is the inverse temperature and $Z$ is related to the time evolution operator $e^{-i \tau H}$ by a Euclidean analytic continuation.
From the geometric point of view, by defining $\tilde \tau = i \tau$ and $ \tilde \xi = e^{\alpha \xi}/\alpha$, the resulting euclidean geometry is
\begin{equation}
  ds^2 = d \tilde \xi^2 + \alpha^2\, \tilde \xi^2\, d \tilde \tau^2,
\end{equation}
which has a conical singularity at the origin unless $\tau \sim \tau + i \, \beta$, where
\begin{equation}
	\label{RNtemperature}
  \beta = \frac1T = \frac{2\pi}{\alpha} = \frac{4 \pi r_+^2}{r_+-r_-}
\end{equation}
and this gives an expression of the temperature in terms of the geometric quantities defining the black hole horizons.

Having now a thermodynamic system for which we defined the energy (given by the mass of the black hole $M$) and a temperature $T$, it is natural to define a (Bekenstein--Hawking) entropy $S_{BH}$, such that, for fixed charges, one fulfills the thermodynamic relation
\begin{equation}
  \frac{dS_{BH}}{dM} = \frac1T.
\end{equation}
In the case at hand, namely the Reissner--Nordstr\"om black hole, integration of the previous equation leads to
\begin{equation}
	\label{RNentropy}
  S_{BH} = \pi \, r_+^2 = \pi \left[M + \sqrt{M^2 - (P^2 + Q^2)}\right]^2.
\end{equation}
The dependence of the entropy on the mass and charge of the black hole is summarized by the geometric quantity $r_+$, the horizon radius, which can be translated to the horizon area, leading to the famous relation
\begin{equation}
	\label{Beckhawking}
  S_{BH} = \frac{A}{4},
\end{equation}
which is also valid for other configurations at the two-derivatives level.
This is a remarkable relation between the thermodynamic properties of a black hole on the one hand and its geometric properties on the other and it is a cornerstone for our understanding of any theory of quantum gravity.
In fact, if we believe that $S_{BH}$ has the meaning of a real entropy, although such a quantity is usually defined in terms of global properties of the system, it contains non trivial information about the microscopic structure of the theory via Boltzmann's relation
\begin{equation}
  S = \log \Omega,
\end{equation}
where $\Omega$ is the total number of microstates of the system for a given energy and fixed total charges.
In detail, the entropy contains information about the total number of microscopic degrees of freedom of the system and in our case a microscopic theory of gravity should explain the black hole entropy in terms of the quantum numbers defining the solution:
\begin{equation}
  S_{BH} = \log \Omega(M,Q,P).
\end{equation}
Explaining this formula is actually one of the biggest problems in theoretical physics.
Given (\ref{Beckhawking}) relating the entropy of a black hole to its horizon area, we can actually see that the typical number of microstates forming a black hole is humongous. 
For instance, the black hole at the centre of our galaxy (Sgr A$^*$) is estimated to have a radius of about $r_+ \sim 7 \cdot 10^9\, Km$ \cite{Ghez:2008ms}, leading to an estimate of $S_{BH} \sim 10^{100}$, and this is just the logarithm of the number of states defining the black hole!
If, on the other hand, we think about a generic black hole solution in GR, we know that the no-hair theorem tells us that a black hole is completely specified by its mass and charges. 
This would mean that, for fixed mass and charges, there is a unique classical state, leading to an expectation of $S = 0$.

It is actually interesting that the laws of black hole mechanics can be put in a one to one relation with the laws of thermodynamics \cite{Townsend:1997ku}:
\begin{itemize}
  \item Zeroth law: the temperature of a black hole $T = \alpha/2 \pi$ is uniform at the horizon;
  \item First law: for quasi static changes the energy (mass) of a black hole changes as
  \begin{equation}
     dM = T dS + \psi dQ + \chi dP + \Omega dJ - g_{ij} \Sigma^i d \phi^j,
  \end{equation}
	where the entropy is identified with the area of the horizon as in (\ref{Beckhawking}), $Q$ and $P$ are the electric and magnetic charges, $J$ is the angular momentum, $\psi$, $\chi$ and $\Omega$ are the associated chemical potentials (namely the electric and magnetic potentials at the horizon and the angular velocity, assumed constant for stationary solutions), $\phi^i$ the values of the scalar fields defining the solution and $\Sigma^i$ the scalar charges;
  \item Second law: the horizon area always increases in time $\Delta A \geq 0$. 
A consequence of this last law is that coalescence processes are possible, while generically no splitting processes are allowed. 
For instance, two Schwarzschild black holes with masses $M_1$ and $M_2$ can form a bigger black hole with mass $M_1+M_2$ because their horizon area is proportional to the square of the corresponding masses and therefore $(M_1+M_2)^2 \geq M_1^2 + M_2^2$.
The inverse process is forbidden by the same argument.
\end{itemize}


Coming back to our example, the Reissner--Nordstr\"om black hole, we can see that the temperature $T$, the entropy $S_{BH}$ and the extremality parameter $c$ are all defined in terms of the characteristic geometric quantities of the solution, namely the radii of the two horizons.
This implies that, by comparing (\ref{RNentropy}), (\ref{RNtemperature}) and (\ref{cdef}), we can express the extremality parameter in terms of the temperature and entropy as
\begin{equation}
  T = \frac{\alpha}{2\pi} = \frac{c}{2S} \quad \Rightarrow\quad  c = 2 \,S\, T.
\end{equation}
Recalling that an extremal configuration is such when the two horizons coincide, i.e.~$c=0$, and that the entropy is non-vanishing whenever there is a non-trivial horizon $S  = \pi\, r_+^2$, we can see that extremality implies vanishing temperature:
\begin{equation}
  {\rm Extremality} \qquad \Leftrightarrow\qquad c = 2\,ST =0 \qquad \Rightarrow\qquad T = 0.
\end{equation}
Extremal black holes are therefore thermodynamically stable. 
They do not radiate.
We will come back to an explanation of this fact momentarily.

The special properties of this kind of black holes is reflected also in the near horizon metric, which now is not given by (\ref{NHnonextremal}) anymore.
Since the warp factor has a double zero, its behaviour close to the horizon is approximated by a quadratic function of $\rho$ rather than linear as in (\ref{warpnonextremal})
\begin{equation}
  e^{2U} = \frac{(r_+-r_-)^2}{r^2} \quad \rightarrow\quad \frac{\rho^2}{r_+^2}.
\end{equation}
The near-horizon metric changes accordingly and, by introducing
\begin{equation}
  z = - \frac{M^2}{\rho},
\end{equation}
we can see that it is given by the product of a 2-dimensional Anti-de Sitter spacetime and a 2-dimensional sphere, both with radius $M = \sqrt{P^2+Q^2}$:
\begin{equation}
  ds^2_{NH} = M^2 \left(\frac{-dt^2+dz^2}{z^2}\right) + M^2 d \Omega^2\,.
\end{equation}
Remember that, using these coordinates, the horizon sits at $z \to - \infty$.
It is interesting to note that this geometry is conformally flat (extremal Reissner--Nordstr\"om solutions are also supersymmetric).

Before proceeding to a more detailed analysis of the differences between extremal and non-extremal black holes, let us pause for a second to make some comments.
From the above discussion we can see that black holes are rather special thermodynamic systems, because they do not satisfy Nernst law, which states that the entropy should vanish (or arrive at a  ``universal constant'' value) as the temperature approaches zero. 
The analog of this law fails in black hole mechanics, because extremal black holes have vanishing temperature, but non-vanishing entropy, $S = \pi \sqrt{P^2 + Q^2}$ in the previous example.
However, there is good reason to believe that ``Nernst theorem'' should not be viewed as a fundamental law of thermodynamics but rather as a property of the density of states near the ground state in the thermodynamic limit, which happens to be valid for commonly studied materials. 
Indeed, examples can be given of ordinary quantum systems that violate the Nernst form of the third law in a manner very similar to the violations of the analog of this law that occur for black holes \cite{Wald:1997qp}.

Another interesting observation follows from rewriting the metric ansatz in an isotropic form:
\begin{equation}
	ds^2 = -H^{-2}(\vec x)\,dt^2 + H^2(\vec x) d\vec x_3^2.
\end{equation}
Once the metric is written in this fashion, the equations of motion for the warp factor can be expressed as 
\begin{equation}
	\triangle_3 H = 0
\end{equation}
and therefore can be solved by generic harmonic functions, which may have more than one centre:
\begin{equation}
  H = 1 + \sum_i \frac{m_i}{|\vec x - \vec x_i|}, \qquad m_i = \sqrt{p_i^2 + q_i^2},
\end{equation}
where $\vec x_i$ denotes the position of the $i$-th centre.
This solution is allowed by the fact that gravitational attraction equals  electromagnetic repulsion for each centre and hence leads to a condition of static neutral equilibrium.
The additive nature of the solution is related to the BPS nature of force-free objects.

Finally, the fact that the near-horizon geometry approaches the product of an Anti-de Sitter spacetime and a sphere is actually a universal behaviour of extremal $p$-branes in $D$ dimensions, whose near horizon geometry is given by AdS$_{p+2} \times S^{D-p-2}$.
Black holes in 4 spacetime dimensions are a simple instance where $p=0$ and $D=4$, but one could also think of different examples like black holes and black strings ($p=0,1$) in $D=5$, dyonic black strings ($p=1$) in $D=6$ and D3-branes in IIB string theory ($p=3$, $D=10$).

\subsection{Extremal vs non-extremal solutions} 
\label{sub:extremal_vs_non_extremal_solutions}

We can now go back to the concept of extremality to discuss an important difference between extremal and non-extremal black holes.
A general ansatz for the metric that satisfies the requirements of describing spherically symmetric, static, asymptotically flat black holes and which encompasses both the extremal and non-extremal solutions is the following:
\begin{equation}
	\label{generalmetric}
  ds^2 = - e^{2U} dt^2 + e^{-2U}\left[\frac{c^4}{\sinh^4(c z)}dz^2 + \frac{c^2}{\sinh^2(cz)}d \Omega^2\right].
\end{equation}
The extremality parameter $c$ was explicitly inserted and the extremal case is recovered by sending $c \to 0$, so that the metric simplifies to
\begin{equation}
	\label{extremalmetric}
  ds^2_{c=0}= - e^{2U} dt^2 + e^{-2U}\left[\frac{dz^2}{z^4} + \frac{1}{z^2}d \Omega^2\right],
\end{equation}
where one can rewrite the factor in brackets using isotropic coordinates as a plain ${\mathbb R}^3$:
\begin{equation}
  \label{extremalisotropic}
  ds^2_{c=0} = - e^{2U} dt^2 + e^{-2U} d \vec x\,{}^2.
\end{equation}
By applying this ansatz to the Reissner--Nordstr\"om case analyzed before, it is easy to realize that the horizon sits at $z \to -\infty$.

However, proper distance from the horizon has to be computed by using appropriate coordinates.
In the non-extremal case, assuming that the horizon area is finite, one gets that the factor in front of the angular variables should remain finite as $z \to -\infty$.
\begin{wrapfigure}{r}{0.4\textwidth}
\begin{center}
	\includegraphics[scale=1.1]{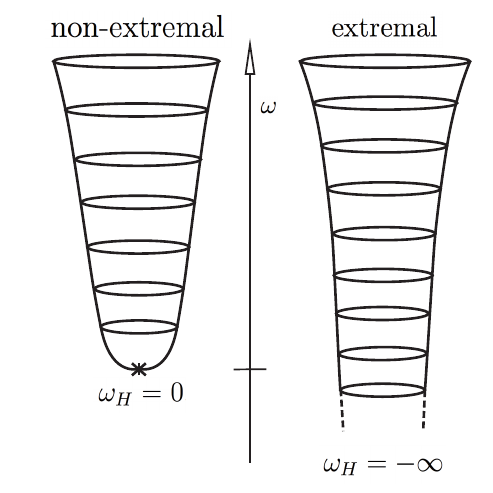} 
\end{center}
\caption{\small \emph{Schematic representation of non-extremal and extremal black hole throats using proper-distance coordinates.}}
 \label{throatplot} 
\end{wrapfigure}
This means that
\begin{equation}
  e^{-2U} \frac{c^2}{\sinh^2(cz)} \quad \stackrel{z\to -\infty}{\longrightarrow} \quad \frac{A}{4\pi} = r_H^2,
\end{equation}
where $r_H$ is the radius of the horizon.
A proper radial coordinate $\omega$ can then be introduced by considering the $g_{zz}$ component of the metric in the same limit:
\begin{equation}
  e^{-2U}\frac{c^4}{\sinh^4(c z)}dz^2 \longrightarrow \frac{A}{4\pi} \,4 c^2 e^{2cz} dz^2 \equiv r_H^2 d \omega^2.
\end{equation}
Distances should then be measured by $\omega = 2 \,e^{c z}$ in units of $r_H$ and the black hole horizon sits at $\omega_H = 0$, at finite proper distance from an arbitrary observer
\begin{equation}
  L = \int_{\omega_H}^{\omega_0} r_H d \omega =r_H\, \omega_0 < \infty.
\end{equation}
On the other hand, having finite area in the extremal case means
\begin{equation}
	\label{asympbehaviour}
  \frac{e^{-2U}}{z^2} \longrightarrow \frac{A}{4\pi} = r_H^2.
\end{equation}
This implies that a new proper radial coordinate can be introduced by identifying 
\begin{equation}
  e^{-2U}\frac{dz^2}{z^4} \longrightarrow \frac{A}{4\pi} \frac{dz^2}{z^2} = r_H^2 \, d \omega^2, 
\end{equation}
which means
\begin{equation}
  \omega = - \log (-z).
\end{equation}
The horizon is now at $\omega_H = - \infty$ at infinite proper distance from any observer
\begin{equation}
  L = \int_{\omega_H}^{\omega_0} r_H d \omega =+ \infty.
\end{equation}
As we will see in a moment, this difference has a crucial impact on the behaviour of scalar fields in this scenario and implies the existence of an attractor mechanism for extremal black hole configurations.
Moreover, the fact that the horizon is at infinite proper distance from any observer justifies also the fact that extremal black holes are thermodynamically stable.
Any radiation emitted by such black hole would be infinitely red-shifted before reaching any observer outside the horizon.



\section{Attractors} 
\label{sec:attractors}

\subsection{Black holes and scalar fields} 
\label{sub:scalars}

When dealing with supergravity theories, as with many other effective theories of fundamental interactions, gravity needs to be coupled to scalar fields, possibly parameterizing a scalar $\sigma$-model and affecting also the couplings of the vector fields (which we consider abelian for the sake of simplicity).
A generic Lagrangian describing the bosonic degrees of freedom of such theories will have the form
\begin{equation}
	\label{generallagrangian}
  e^{-1}{\cal L} = R - \frac12\, g_{ij}(\phi)\,\partial_\mu \phi^i \partial^\mu \phi^j + \frac14\, {\cal I}_{\Lambda \Sigma}(\phi) F^\Lambda_{\mu\nu}F^{\Sigma\,\mu\nu} + \frac14\, {\cal R}_{\Lambda \Sigma}(\phi) \frac{\epsilon^{\mu\nu\rho\sigma}}{2\sqrt{-g}}F_{\mu\nu}^\Lambda F_{\rho\sigma}^\Sigma,
\end{equation}
where $g_{ij}(\phi)$ is the metric of the scalar $\sigma$-model, $\cal I$ is definite negative and describes the gauge kinetic couplings, ${\cal R}$ is the generalization of the $\theta$-angle terms in the presence of many scalar and vector fields and we assume that there is no scalar potential for the time being.
We are still interested in finding single centre, static, spherically symmetric, charged and asymptotically flat black hole solutions and therefore we keep the metric ansatz (\ref{generalmetric}) and the request that the integral of the vector field strengths and their duals on a sphere at infinity gives the electric and magnetic charges of the solution:
\begin{equation}
	\label{chargesint}
  \frac{1}{4\pi} \int_{S^2} F^\Lambda = p^\Lambda\,, \qquad \qquad   \frac{1}{4\pi} \int_{S^2} G_{\Lambda} = q_{\Lambda}.
\end{equation}
Now, however, we need to introduce a new definition of the dual field strengths $G_{\Lambda}$, because the values of the gauge couplings and the charges will be affected by the values of the scalar fields appearing in the functions ${\cal R}$ and $\cal I$.

In the setup considered in the previous section, the magnetic and electric charges are associated to the 2-forms appearing in the Bianchi identities and in the equations of motion of the electromagnetic theory, respectively.
These 2-forms are also related between them by the known electromagnetic duality $F \leftrightarrow \star F$.
In a general setup, like the one considered here, electric--magnetic duality can be extended to a new group of duality transformations that leaves invariant Bianchi identities and equations of motion \cite{Gaillard:1981rj}.
If we focus on the part of the Lagrangian involving the gauge field-strengths
\begin{equation}
  S_{EM} = \int \left[{\cal I}_{\Lambda \Sigma}F^\Lambda \wedge \star F^\Sigma - {\cal R}_{\Lambda \Sigma}F^\Lambda \wedge F^\Sigma\right],
\end{equation}
we can deduce that the Bianchi identities and equations of motion form a set, from which we can define the dual field strengths $G_{\Lambda}$:
\begin{equation}
  \left\{\begin{array}{l}
   dF^\Lambda = 0\,, \\[2mm]
   d G_{\Lambda}=d\left({\cal R}_{\Lambda \Sigma}F^\Sigma - {\cal I}_{\Lambda \Sigma} \star F^\Sigma\right) = 0\,.
  \end{array}\right.
\end{equation}
It is obvious, that for any constant matrix $S$ we can rotate the original field strengths $F^\Lambda$ and the dual ones $G_{\Lambda}$ between them, leaving the full set of Bianchi identities and equations of motion invariant:
\begin{equation}
  \left(\begin{array}{c}
  F \\[2mm] G
  \end{array}\right) \to 
  \left(\begin{array}{c}
  F^\prime \\[2mm] G^\prime
  \end{array}\right)=
  S \left(\begin{array}{c}
  F \\[2mm] G
  \end{array}\right).
\end{equation}
However, the requirement that also the definition of the dual field-strengths $G_{\Lambda} \equiv - \frac{\delta {\cal L}}{\delta F^\Lambda}$ remains invariant constrains the duality transformation to be part of the symplectic group $S \in $ Sp($2 n_V, {\mathbb R}$), where $n_V$ is the total number of abelian vector fields in the theory and the symmetry transformations are continuous at the classical level.
Moreover additional matter couplings, like the ones considered in (\ref{generallagrangian}), may reduce it to $G \subset$ Sp($2 n_V, {\mathbb R})$. 
$G$ is called the U-duality group of the theory.
An important result of \cite{Gaillard:1981rj} is that the stress energy tensor and hence the Einstein equations of motion following from rather general interactions between the various fields are invariant under such transformations.
This means that by using the duality group we can map solutions of the Bianchi identities and of the equations of motion to new solutions of the same set of equations, leaving the metric untouched.
In particular, we can map charged black hole solutions with different charges and scalar fields between them without changing the metric and hence a crucial property like the area of the horizon.

Before proceeding, let us note that by performing such duality transformations the Lagrangian does not necessarily remain invariant.
U-duality transformations are symmetry transformations by which the equations of motion and the Bianchi identities are mixed among themselves linearly and this may require changes in the Lagrangian originating them.

Since $(F^\Lambda, G_{\Lambda})$ form a symplectic vector of closed 2-forms, we could introduce explicitly the corresponding potentials $(A_\mu^\Lambda,A_{\mu\Lambda})$, though obviously not both at the same time.
Given the request (\ref{chargesint}), the vector potentials should have a restricted form such that the integrals provide the correct electric and magnetic charges.
In particular, solving $dF^\Lambda = 0$ and respecting the request of finding solutions with spherical symmetry, we can introduce
\begin{equation}
  A^\Lambda = \chi^\Lambda(r) dt - p^\Lambda \cos \theta\, d \phi,
\end{equation}
where $\chi^\Lambda$ are the electric potentials, so that $F^\Lambda = d A^\Lambda$.
In the same fashion we can also introduce the dual potentials 
\begin{equation}
  A_{\Lambda} = \psi_{\Lambda}(r) dt - q_{\Lambda} \cos \theta\, d \phi,
\end{equation}
where $\psi_{\Lambda}$ are the magnetic potentials, so that $G_{\Lambda} = d A_{\Lambda}$.
In the original action (\ref{generallagrangian}) only the first appears.
Actually, we can see that $\chi^\Lambda$ appears in the action only under derivatives and therefore we can integrate it out.
In fact, from the $\chi^\Lambda$ equations of motion one gets
\begin{equation}
	\label{chiprime}
  \chi^\Lambda{}^\prime = e^{2U} {\cal I}^{-1\,\Lambda \Sigma} \left(q_{\Sigma} - {\cal R}_{\Sigma \Gamma}\,p^\Gamma\right),
\end{equation}
which is also the correct relation needed to fulfill the duality relation by which the definition of $G_{\Lambda}$ follows from the one of $F^\Lambda$.

A simple strategy to find black hole solutions in this framework is to use the fact that the problem is spherically symmetric, so that all relevant quantities depend only on the radial variable,  $\phi^i = \phi^i(r)$, $U= U(r)$, etc., and reduce the system to 1 dimension. 
By using (\ref{chiprime}) and by integrating out formally $\theta$, $\phi$ and $t$ one gets the effective 1-dimensional action
\begin{equation}
	\label{1dL}
  L_{1d} = (U')^2 + \frac12\,g_{ij} \phi^i{}'\phi^j{}' + e^{2U} V_{BH} - c^2,
\end{equation}
where
\begin{equation}
	\label{VBH}
  V_{BH} = -\frac12 Q^T {\cal M} Q,
\end{equation}
with
\begin{equation}
  {\cal M} = \left(\begin{array}{cc}
  I + R I^{-1} R &  - R I^{-1} \\
  -I^{-1}R & I^{-1}
  \end{array}\right)
\end{equation}
and
\begin{equation}
  Q =  \left(\begin{array}{c}
  p^\Lambda \\[2mm] q_{\Lambda}
  \end{array}\right).
\end{equation}
The kinetic term for the warp factor and the overall constant $c^2$ come from the reduction of the Einstein kinetic term.
The black hole potential $V_{BH}$ comes from the reduction of the kinetic term and $\theta$-angle terms of the vector fields, after dualization of the electric potential using (\ref{chiprime}).
The resulting problem is a 1-dimensional mechanical system of $n+1$ scalars in the presence of a potential $V_{BH}$ depending on a number of parameters equal to the total number of non-vanishing electric and magnetic charges.

Since we have made an ansatz on the metric, we should take into account the possibility that the equations of motion of this reduced system do not solve also the equations of motion of the original one, because we are essentially looking at constrained variations of the original system.
This is actually the case and it means that in order to obtain solutions from the equations of motion coming from this Lagrangian that are equivalent to the original ones, we need to supplement the 1-dimensional Lagrangian (\ref{1dL}) with the constraint
\begin{equation}
   (U')^2 +\frac12\, g_{ij} \phi^i{}'\phi^j{}' = e^{2U} V_{BH} + c^2.
\end{equation}

For completeness, we provide here the equations of motion:
\begin{eqnarray}
	\label{earpeom}
  U'' &=& e^{2U} V_{BH},\\[2mm]
\label{scalareom}
\phi^i{}'' + \Gamma_{jk}{}^i \phi^j{}' \phi^k{}'  &=&  e^{2U} g^{ij} \partial_j V_{BH}.
\end{eqnarray}


\subsection{General features of the attractor mechanism} 
\label{sub:generic_features_of_the_attractor_mechanism}

We have seen that black holes in generic supergravity theories will depend on scalar fields.
However, extremal black holes have the special property that the horizon quantities loose all the information about them.
This is true independently of the fact that the solution preserves any supersymmetry or not.
The horizon is in fact an attractor point \cite{Ferrara:1995ih,Ferrara:1996dd,Ferrara:1996um,Strominger:1996kf}: scalar fields, independently of their value at spatial infinity, flow to a fixed point given in terms of the charges of the solution at the horizon.
Recalling that the entropy of black holes is given by the area of the horizon, this attractive behaviour for the scalar fields implies that for extremal black holes the entropy is a topological quantity, given in terms of quantized charges and therefore it does not depend on continuous parameters, which is a very appealing feature in order to have the chance to provide a microscopic explanation for the resulting number.

The main reason at the base of the attractor mechanism is the fact that for extremal black holes the horizon is at an infinite proper distance from any observer \cite{Ferrara:1997tw}.
This means that while moving along the infinite throat leading to the horizon, scalar fields lose memory of the initial conditions.
This is an obvious outcome of the request of having regular solutions.
In fact, regularity of the scalar fields at the horizon implies that their derivative should vanish while approaching the horizon
\begin{equation}
  \phi^i{}' \ \stackrel{z \to -\infty}{\longrightarrow} \ 0. 
\end{equation}
On the other hand non-extremal black holes have $\omega_H = 0$ and hence scalar fields do not have time to blow up even for a non-trivial, but finite, first derivative along the radial direction.

Extremal black holes are therefore described by trajectories in the moduli space with a fixed point reached when the proper radial parameter $\omega \to - \infty$.
The fixed point is an attractor of the system.
Since infinitely far away from the black hole ($\omega \to + \infty$) the geometry approaches that of 4-dimensional Minkowski spacetime and at the horizon ($\omega \to - \infty$) it approaches the product AdS$_2 \times S^2$, we can see that extremal black holes can also be thought of as solitons interpolating between two different vacua of the theory.
We will come back to this picture later on to justify the description of such solutions in terms of first-order differential equations.

\begin{figure}
	[htb] \centerline{  
	\includegraphics[scale=1.2]{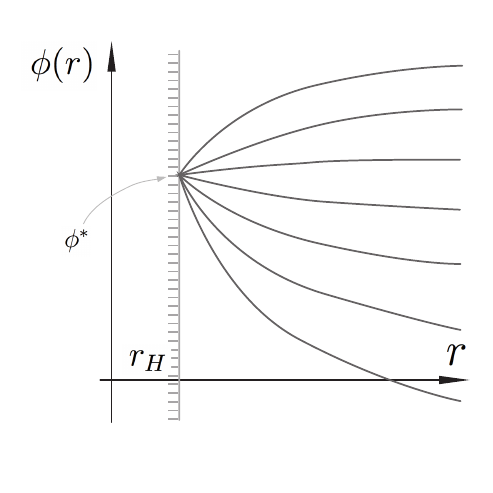} \hspace{2cm}
	\includegraphics[scale=1.2]{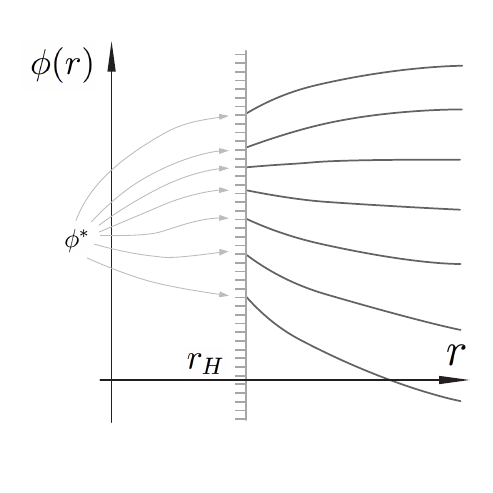} } \caption{Generic behaviour of a scalar field in the case of an extremal black hole (left) and of a non-extremal one (right). In the first case the scalar field $\phi$ runs towards the same value at the horizon $\phi^*$, no matter what was its value at infinity. In the second case the scalar stops at different points in moduli space depending on its asymptotic value.} \label{attractorsfig} 
\end{figure}

{}From the equations of motion of the scalar fields (\ref{scalareom}), fixed scalars at the horizon imply that the moduli reached a critical point of the black hole potential:
\begin{equation}
	\label{minimum}
  \partial_i V_{BH}(\phi^{i*}, q,p) = 0.
\end{equation}
This is actually an intrinsic characterization of the horizon for extremal black holes.
Extremization of the scalar potential, means that the scalar fields at the horizon take the value $\phi^{i*}$ such that the minimization condition (\ref{minimum}) is satisfied.
Since the only parameters appearing in the minimization condition are the black hole charges, the resulting attractor values of the moduli fields are also going to be given in terms of the same charges 
\begin{equation}
	\phi^{i*} = \phi^{i*}(q,p).
\end{equation}
In turn, this implies that at the horizon the value of the scalar potential does not depend anymore on the values of the scalars at infinity, but only on the charges:
\begin{equation}
	V_{BH}^* = V_{BH}(\phi^{i*}(q,p),q,p).
\end{equation}
At this point we can also solve the equation for the warp factor (\ref{earpeom}) close to the attractor point, which gives that
\begin{equation}
  U \to - \log\left(\sqrt{V_{BH}^*}z\right). 
\end{equation}
This implies that the metric approaches that of AdS$_2 \times S^2$ as expected, with a characteristic horizon radius given by $r_H = \sqrt{V_{BH}^*}$.
This in turn implies that the entropy of extremal black holes can be expressed in terms of the value of the black hole potential at the critical point:
\begin{equation}
  \label{SBHextr}
  S_{BH} = \frac{A}{4}=\pi \, V_{BH}^*(q,p).
\end{equation}
Since the black hole potential depends only on the quantized charges, because of the attractor mechanism, also the entropy of extremal black holes will depend on the same quantized parameters and all the possible dependence on the value of the moduli fields (which still characterize the full solution) is lost.

The contrast becomes even more clear if we compare (\ref{SBHextr}) with the corresponding expression for non-extremal solutions, where the area formula is valid for a radius of the horizon sitting at the larger value between
\begin{equation}
  r_\pm = M \pm \sqrt{M^2 - V_{BH}(\phi_{\infty}, p,q) + \frac12\, g_{ij}(\phi_\infty)\Sigma^i \Sigma^j}.
\end{equation}
Not only this expression depends on the value of the scalar fields at infinity, but also on the scalar charges, which vanish only for solutions where the scalars remain constant \cite{Gibbons:1996af}.

\begin{figure}
	[htb] 
	\center{  
	\includegraphics[scale=.13]{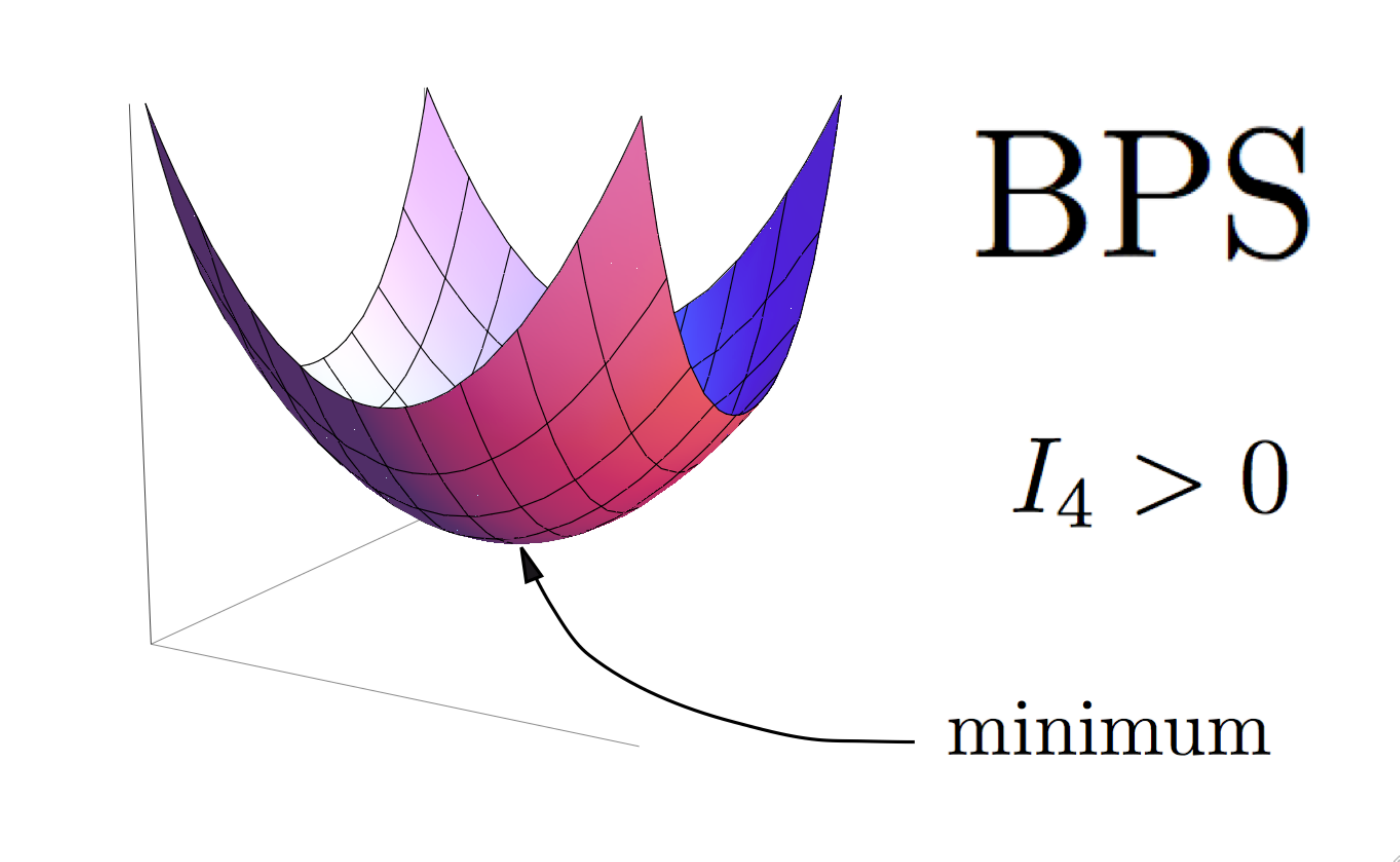}\hfill
	\includegraphics[scale=.14]{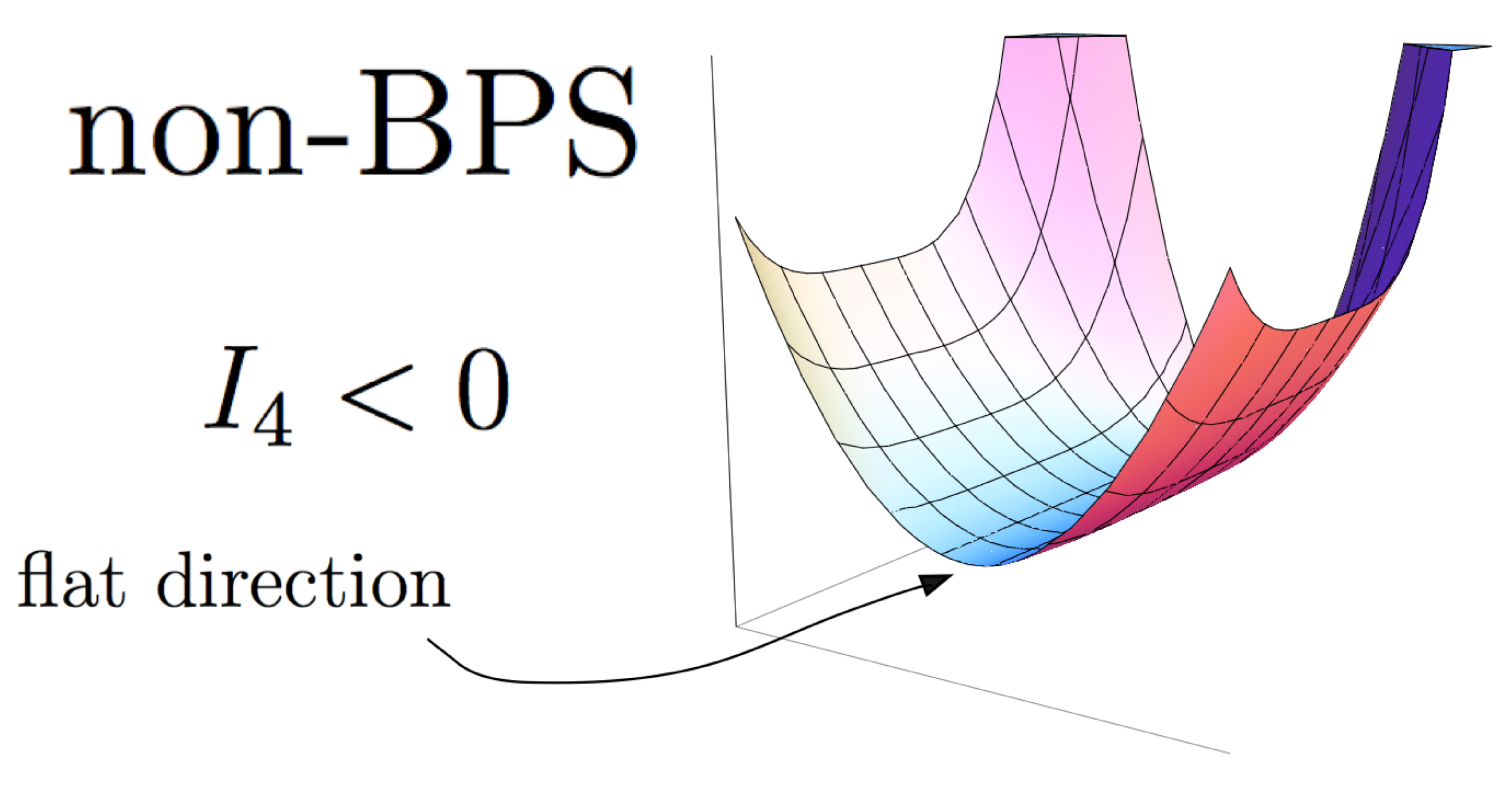} 
	}    
	\caption{\small \emph{Examples of black hole potentials for various values of the quartic invariants.}} \label{N8potentials} 
\end{figure}

An interesting outcome of this analysis is that  extremal black hole solutions are completely specified by the black hole potential.
In particular different kind of attractors will be characterized by different types of potentials.
Still, while $V_{BH}$ depends on the theory under investigation, the general features of the attractor mechanism are universal.

As an example, consider the most constrained supergravity theory in 4 dimensions: maximally supersymmetric ($N=8$) supergravity.
This theory has a fixed matter content, which is all contained in the gravity multiplet.
Among other fields, the gravity multiplet contains 28 vector fields, leading to 56 charges, and 70 scalar fields parameterizing the scalar manifold E$_{7(7)}$/SU(8).
The black hole potential depends on the detailed choice of the 28 electric and 28 magnetic charges, but, given the restrictive form of the scalar manifold and of the invariances of the theory, one can distinguish three main classes of solutions.
These are related to the value of a special E$_{7(7)}$ invariant, which is quartic in the charges $I_4(p,q)$ \cite{Kallosh:1996uy,Ferrara:1997uz}.
Whenever $I_4 >0$ the scalar potential has a minimum and the corresponding black hole solutions preserve some supersymmetry.
If $I_4<0$ the solutions are non-supersymmetric.
Finally, in the special instance where the quartic invariant vanishes, the warp factor at the horizon vanishes.
This implies that the corresponding classical geometry is singular and various orbits can be further distinguished by the values of derivatives of $I_4$.
\begin{wrapfigure}{r}{0.4\textwidth} 
\begin{center}
	\includegraphics[scale=.075]{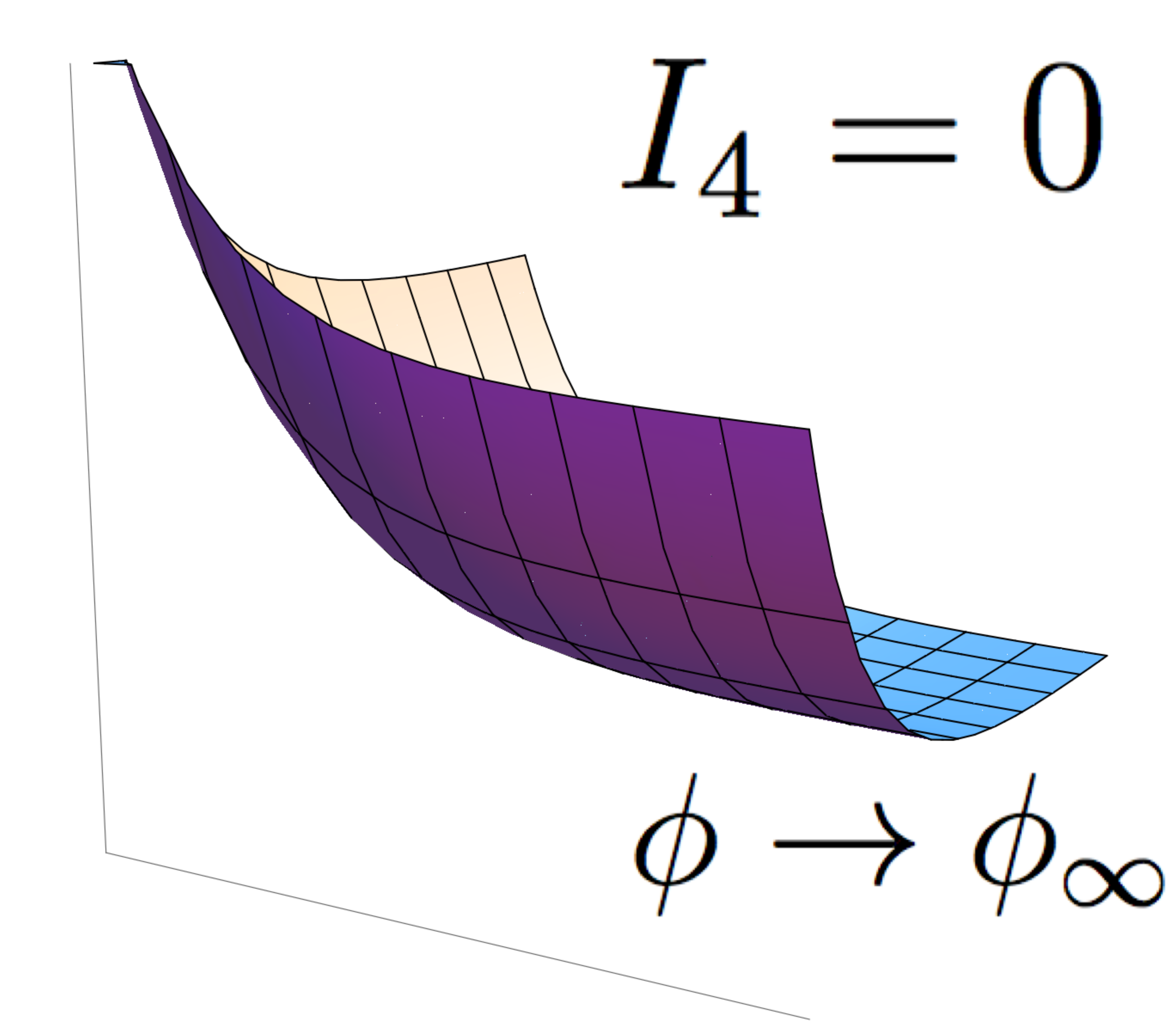} 
\end{center}
\caption{\small \emph{Runaway behaviour for small black holes}} \label{smallBHpotential} 
\end{wrapfigure}
However, higher-order corrections in the curvature terms modify the equations of motion in a way such that a horizon is developed, with a characteristic radius of the order of the typical scale of the correction terms.
For this reason the corresponding black holes are called small black holes.

In generic $N=2$ theories supersymmetric configurations are always minima, while non-BPS ones have flat directions at the attractor point the potential (actually these flat directions are generically given by expectation values of scalar fields that do not appear in the scalar potential at all) \cite{Ferrara:2007tu}.
For $N>2$ also supersymmetric attractors may have a non-trivial moduli space.

One important lesson that can be learned from this analysis is that while supersymmetry always implies extremality, the opposite is not true.
In fact the supersymmetry condition is achieved when the mass of the BPS object equals a certain value defined by its charges, however, for a given charge configuration the BPS bound may never be reached and hence for such configurations the object with the minimal mass will still be extremal, though non supersymmetric.



\section{Glancing through Special K\"ahler geometry} 
\label{sec:special_k"ahler_geometry}

As discussed in the previous section, the black hole potential containing the necessary information to describe extremal black holes depends on the detail of the model under investigation.
General features of these solutions can anyway be obtained independently on such details.
Although in the following we will try to give some general arguments about the properties of single and multi-centre extremal black holes in supergravity theories, it is better to fix a specific framework, so that we can provide explicit examples along with the general arguments.
For this reason we now provide a brief \emph{intermezzo} with some elementary facts about Special K\"ahler geometry, which is the geometric structure underlying the vector multiplet scalar $\sigma$-model in ${\cal N} =2$ theories in 4 dimensions.

We will not give an exhaustive review of this topic, but rather focus on some minimal ingredients  necessary for our following discussion.
An interested reader can find more details on the many geometric identities and on the relation with String Theory in \cite{Ceresole:1995ca} and in references therein.
There are three main types of multiplets in ${\cal N} = 2$ supergravity: gravity, vector and hyper-multiplets
\begin{equation}
	\begin{array}{ccccc}
	\left(\begin{array}{c}
	g_{\mu\nu}\\[2mm]
	\psi_\mu^A \\[2mm]
	A_\mu^0
	\end{array}\right),& \hspace{5mm}&
	\left(\begin{array}{c}
	A_{\mu}^i\\[2mm]
 	\lambda_A^i\\[2mm]
	z^i
	\end{array}\right), &\hspace{5mm} &
	\left(\begin{array}{c}
	\zeta^\alpha\\[2mm]
 	q^u
	\end{array}\right), \\
	&& \\
	{\rm gravity}, & &n_V\ {\rm vector} &	&n_H\ {\rm hyper} \\
&	& {\rm  multiplets},& & {\rm  multiplets}.
\end{array}
\end{equation}
In a generic interacting ungauged theory, the number of vector fields is determined by the number of vector multiplets $n_V$ with the addition of the vector field sitting in the gravity multiplet, named graviphoton.
Scalar fields sit in both vector ($2 n_V$ real fields) and hypermultiplets ($4 n_H$ real fields).
The self-interactions between these fields can be described by a factorized $\sigma$-model given by the product of a Special K\"ahler manifold for the scalars in the vector multiplets and a Quaternionic K\"ahler manifold for the scalars in the hypermultiplets:
\begin{equation}
	{\cal M}_{\rm scalar} = {\cal M}_{SK} \otimes {\cal M}_{QK}.
\end{equation}
The different structure between the two manifolds has to do with the way the U(2) $R$-symmetry group of ${\cal N} = 2$ theories acts on the fields of the two multiplets.
Just like the $R$-symmetry group factorizes U(2) $=$ U(1) $\times$ SU(2), so does the scalar manifold.

The scalars in the vector multiplets define the gauge kinetic functions $\cal I$ and $\cal R$, while the hyperscalars do not enter into their definition.
For this reason the scalars of the hypermultiplets do not appear in the black hole potential $V_{BH}$ and hence do not participate at the definition of the tree level solutions.
Hence, we will set them to zero for the time being.

A Special-K\"ahler (SK) manifold can be parameterized by $n_V$ complex scalar fields, the scalars appearing in the vector multiplets.
However, SK manifolds have an intrinsic projective nature, related to the fact that there are $n_V+1$ vectors that appear in a supergravity theory that can mix between them.
This means that one could use $n_V+1$ projective coordinates $X^\Lambda(z)$, $\Lambda = 0,1,\ldots, n_V$, which are holomorphic sections of the complex line bundle associated to the scalar manifold (whose principal bundle is related to the U(1) factor in the $R$-symmetry group). 
Actually, just like for any set of $n_V +1$ vector fields we can define a set of $n_V +1$ duals according to the procedure outlined previously and symplectic duality transformations mix them, a SK manifold can be specified in terms of $2 n_V +2$ sections, which form a symplectic vector $(X^\Lambda, F_{\Lambda})$.
Its K\"ahler potential is then defined via these sections as
\begin{equation}
	\label{Kpot}
	K = - \log i\, \left(\bar X^\Lambda F_{\Lambda} - X^\Lambda \bar F_{\Lambda}\right) = - \log i \langle \Omega, \overline \Omega \rangle,
\end{equation}
where
\begin{equation}
	\langle A, B \rangle = A^T \left(\begin{array}{cc}
	0 & -1 \\ 1 & 0
	\end{array}\right) B.
\end{equation}
Summarizing a \emph{SK manifold is a K\"ahler manifold endowed with both a projective and a symplectic structure.}

Obviously the $(X^\Lambda, F_{\Lambda})$ sections and consequently the K\"ahler potential, is only defined locally.
This means that, given two patches covering the scalar manifold $U_{\alpha}$ and $U_{\beta}$, the sections in their non-trivial intersection can be related by a symplectic and holomorphic transformation 
\begin{equation}
	\left(\begin{array}{c}
	X \\ F
	\end{array}\right)_{\alpha} = S_{\alpha \beta} \, e^{h_{\alpha \beta}(z)} \left(\begin{array}{c}
	X \\ F
	\end{array}\right)_{\beta},
\end{equation}
where $S_{\alpha \beta} \in {\rm Sp}(2 n_V + 2, {\mathbb R})$ is constant. 
This implies a K\"ahler transformation on the K\"ahler potential
\begin{equation}
	K_{\alpha} \to K_\beta + h_{\alpha \beta} + \bar h_{\alpha \beta}.
\end{equation}

The projective nature becomes manifest in the fact that there is always a choice of the sections so that normal coordinates can be defined
\begin{equation}
	t^i = \frac{X^i}{X^0}.
\end{equation}
In such frames, the dual sections $F_{\Lambda}(z)$ can be derived from a prepotential $F(X)$, such that $F(\lambda X) = \lambda^2 F(X)$.
We should stress that, on the other hand, generically there are frames in which such a prepotential does not exist.

The geometric structure of SK geometry fixes completely all the other couplings, among which the vector kinetic terms, which can be given in terms of a function ${\cal N}_{\Lambda \Sigma}$, with the property
\begin{equation}
 F_{\Lambda} = {\cal N}_{\Lambda \Sigma} X^\Sigma.
\end{equation}
The gauge kinetic couplings are the real and imaginary parts of this complex matrix:
\begin{equation}
	R_{\Lambda \Sigma} = {\rm Re}\, {\cal N}_{\Lambda \Sigma}, \qquad I_{\Lambda \Sigma} = {\rm Im}\, {\cal N}_{\Lambda \Sigma}.
\end{equation}

\subsection{Examples} 
\label{sub:examples}

Before proceeding further, we give here a couple of interesting examples, which will be used in the following.

The first example is one of the simplest SK manifold one could think of: a manifold with a single scalar field parameterizing SU(1,1)/U(1).
In a frame where a prepotential exists, it is defined as
\begin{equation}
	F = - i X^0 X^1,
\end{equation}
which implies that $F_0 = -i X^1$ and $F_1 = -i X^0$.
In such a frame we can also define a normal coordinate $z = X^1/X^0$ and, with the gauge choice $X^0 =1$, we can write the holomorphic sections and the K\"ahler potential as
\begin{equation}
	\Omega = \left(\begin{array}{c}
	1 \\ z \\ -i z \\ -i
	\end{array}\right), \qquad K = - \log 2 (z+\bar z).
\end{equation}
The $z$ modulus is constrained, because its real part must be positive in order for the K\"ahler potential to be well defined.

A second simple example is the so-called STU model.
This is a scalar manifold corresponding to [SU(1,1)/U(1)]$^3$.
The prepotential is
\begin{equation}
	F = \frac{X^1 X^2 X^3}{X^0}
\end{equation}
and the sections and K\"ahler potential can be written as
\begin{equation}
	\Omega = \left(\begin{array}{c}
	1 \\ s\\t\\u \\ -s t u \\ tu \\ su \\st
	\end{array}\right), \qquad K = - \log [-i(s-\bar s)(t - \bar t)(u - \bar u)],
\end{equation}
in a basis where $X^0 =1$.
It is interesting to point out that the metric of such manifold factorizes.


\subsection{String Theory origin} 
\label{sub:string_theory_origin}

Supergravity theories with ${\cal N} = 2$ supersymmetry in 4 dimensions can be obtained in various ways.
The main path is to consider type II theories in 10 dimensions on Calabi--Yau threefolds.
For instance, type IIB supergravity on a Calabi--Yau manifold $Y_6$ has $2(n_V + 1)$ 3-cycles in $H_3(Y_6,{\mathbb R})$, with $n_V = h_{(2,1)}$, that lead to $n_V$ vector multiplets in the effective theory by reduction of the Ramond--Ramond 4-form of type IIB on them.
The scalar fields in the corresponding SK manifold parameterize the space of complex structure deformations of the internal manifold.
The holomorphic sections we introduced previously can be introduced by considering the periods of the holomorphic 3-form $\Omega$ (with a suggestive abuse of notation) on the symplectic basis $(A_{\Lambda},B^\Lambda)$ of $H_3(Y_6,{\mathbb R})$:
\begin{equation}
	X^\Lambda = \int_{A_{\Lambda}} \Omega, \qquad F_{\Lambda} = \int_{B^\Lambda} \Omega.
\end{equation}
The corresponding K\"ahler potential can be obtained as
\begin{equation}
	K = - \log i \int_{CY} \Omega \wedge \overline \Omega,
\end{equation}
where the analogy between this expression, given in terms of the wedge product of the holomorphic 3-form $\Omega$, and (\ref{Kpot}), given in terms of the symplectic product of the sections $\Omega$, is now evident.

By calling $(\alpha_{\Lambda}, \beta^\Lambda)$ the basis of harmonic 3-forms on $Y_6$, the vector fields arise in the expansion of
\begin{equation}
	F_5 = F^\Sigma \wedge \alpha_{\Sigma} - G_{\Sigma} \wedge \beta^\Sigma,
\end{equation}
where the duality relation between $G_{\Lambda}$ and $F^\Lambda$ follows from the self-duality property of $F_5 = * F_5$.
This expression is also telling us that the black hole charges in 4 dimensions correspond to charges of $F_5$ integrated over the product of a 2-sphere and the 3-cycles of $Y_6$.
This means that black holes can be viewed as the superposition of D3-branes wrapping different 3-cycles of $Y_6$, hence giving a hint on the route one needs to follow to explain the microscopic origin of the entropy of such configurations.

Since any Calabi--Yau manifold has at least a non-trivial 3-cycle associated to the holomorphic form $\Omega$, we can see why there is always at least one vector field in the corresponding ${\cal N} =2$ effective theory, which appears in the gravity multiplet.
The K\"ahler structure deformations are described by the hypermultiplet scalar fields.

Reductions of type IIA supergravity on a Calabi--Yau manifold are similar to the ones just described but with the role of complex and K\"ahler structure reversed: $\Omega \leftrightarrow J$.
In particular, the vector-multiplet moduli space describes the complexified K\"ahler structure of $Y_6$.
If $J_c \equiv B + i\, J$, $C_i$ is a basis of $H_{(1,1)}(Y_6,{\mathbb R})$ and $D^i$ is the dual basis of $H_{(2,2)}(Y_6,{\mathbb R})$,
\begin{equation}
	\frac{X^i}{X^0} = \int_{C_i} J_c, \qquad \frac{F_i}{F_0} = \int_{D^i} J_c \wedge J_c.
\end{equation}
The K\"ahler potential is now
\begin{equation}
	K = - \log\left[\frac43\, \int_{CY} J \wedge J \wedge J\right].
\end{equation}
Vector fields generate from the 1, 3, 5 and 7-form potentials of type IIA expanded on the basis of harmonic 0,2,4 and 6-forms respectively.
This means that the associated charges come from wrapped D0, D2, D4 and D6-branes.



\section{Flow equations for BPS and non-BPS attractors} 
\label{sec:flow_equations}

In this section we are going to show that extremal black holes admit a first order description, no matter whether they are supersymmetric or not.
For the sake of simplicity and in order to be specific, we will constrain our discussion to models within ${\cal N} =2$ supergravity, but the results hold for more general theories.
This presentation follows mainly \cite{Ceresole:2007wx}, where the result was first derived, but expanding on the reasoning justifying and explaining it.

As we saw previously, critical points of the black hole potential define extremal black hole configurations and the same potential plays an essential role in the attractor mechanism.
For ${\cal N} = 2$ theories the potential is
\begin{equation}
	\label{VN2}
	V_{BH} = |Z|^2 + 4 g^{i\bar \jmath} \partial_i |Z| \overline\partial_{\bar \jmath} |Z|,
\end{equation}
where 
\begin{equation}
	Z = e^{-K/2} \left(X^\Lambda q_{\Lambda} - p^\Lambda F_{\Lambda}\right) = e^{-K/2} \langle \Omega, Q\rangle
\end{equation}
is the central charge of the ${\cal N} = 2$ supersymmetry algebra.

Extremal black holes are solutions of the equations of motion derived from the effective 1-dimensional lagrangian
\begin{equation}
	\label{N2lagrangian}
	{\cal L} = (U')^2 + g_{i \bar \jmath} z^i{}' \bar z^{\bar \jmath}{}' + e^{2 U}\left( |Z|^2 + 4 g^{i\bar \jmath} \partial_i |Z| \overline\partial_{\bar \jmath} |Z|\right),
\end{equation}
also satisfying the constraint
\begin{equation}
	\label{N2constraint}
	H=0 \qquad \Leftrightarrow \qquad (U')^2 + g_{i \bar \jmath} z^i{}' \bar z^{\bar \jmath}{}' = e^{2 U}\left( |Z|^2 + 4 g^{i\bar \jmath} \partial_i |Z| \overline\partial_{\bar \jmath} |Z|\right)
\end{equation}
and where the scalar fields reach a critical point of the potential.
The generic equations that need to be satisfied are second-order equations.
We will now show that we can actually further reduce the system to first-order ordinary differential equations.

\subsection{Supersymmetric attractors} 
\label{sub:susy_attractors}

The Hamiltonian constraint (\ref{N2constraint}) is an equality between two different sums of squares weighted with the same positive definite metric $g_{i \bar \jmath}$.
A natural solution is given by matching each term on the left hand side with the corresponding term on the right hand side as $	U' = \pm e^U |Z|$ and $	z^i{}' = \pm 2\, e^U g^{i\bar \jmath} |Z|$, for an arbitrary choice of sign in both equations.
Although surprising at first sight, it is a straightforward exercise to show that such a solution of the constraint equation is also a solution of the equations of motion coming from (\ref{N2lagrangian}), provided the same sign is chosen in the flow equations.
We therefore reduced the system of second-order equations of motion and the quadratic constraint to a system of first order ordinary differential equations driven by the absolute value of the central charge $|Z|$.

The flow equation for the warp factor can also be rewritten as $(e^{-U})' = \mp |Z|$ and should be increasing along the flow, because its value is going to be 1 at infinity and becomes proportional to $|z|$ when approaching the horizon (see the discussion around (\ref{asympbehaviour})).
This means that only the lower sign is acceptable in order to generate regular black hole solutions and hence black holes can be described by the following set of flow equations:
\begin{equation}
	\label{susyflow}
	\left\{\begin{array}{l}
	U' = - e^U |Z|,\\[3mm]
	z^i{}' = - 2\, e^U g^{i\bar \jmath} \overline \partial_{\bar \jmath}|Z|.
	\end{array}\right.
\end{equation}
Having first-order equations rather than second-order, may be a sign of supersymmetry and in fact this is the case at hand.
The mass of the black holes generated by (\ref{susyflow}) is
\begin{equation}
	M_{ADM} = |Z|_{\infty},
\end{equation}
which means that they are extremal configurations at the threshold of the supersymmetric bound $M \geq |Z|$.
In fact, by analyzing the gravitino and gaugino supersymmetry transformations one finds that, after imposing a suitable projector on the supersymmetry parameter, the first flow equation is equivalent to $\delta \psi_\mu^A = 0$, while the second satisfies $\delta \lambda^i_A = 0$.
Actually, the scalar equation coming from the supersymmetry variation of the gauginos is 
\begin{equation}
	z^i{}' = -  e^{U-i \alpha} g^{i\bar \jmath} \overline D_{\bar \jmath} \overline Z,
\end{equation} 
where $\alpha$ is a phase factor appearing in the projector, identified with the phase of the central charge.
Full equivalence with (\ref{susyflow}) can be established by realizing that also the phase obeys a first order equation coming from the consistency of the supersymmetry conditions
\begin{equation}
	\alpha' + Q = 0,
\end{equation}
where $Q$ is the composite K\"ahler connection $Q =$ Im $z^i{}' \partial_i K$, and that this equation is identically satisfied once the flow equations (\ref{susyflow}) are fulfilled.
This is an obvious consequence of the fact that $\alpha$ is not a new independent degree of freedom.

Inspection of (\ref{susyflow}) also shows that the central charge $|Z|$ determines completely the solution and that, no matter what is the value of the scalar fields at infinity, the flow stops where the central charged is minimized: 
\begin{equation}
	\label{susyattractor}
	\partial_i |Z|_* = 0 \quad \Leftrightarrow \quad z^i{}' = 0.
\end{equation}
As we could expect, such a critical point of the central charge is also a critical point of the full black hole potential $V_{BH}$:
\begin{equation}
	\partial_i V_{BH} = |Z| \partial_i |Z| + \partial_i \partial_j |Z| \bar \partial^j |Z| + \partial_j |Z|   \partial_i\bar \partial^j |Z|  = 0
\end{equation}
and therefore a generic flow that reaches such a critical point describes a supersymmetric extremal black hole.

As expected, the attractor conditions (\ref{susyattractor}) fix the values of the scalar fields in terms of the asymptotic charges of the solution $z^{i*} = z^{i*}(p,q)$ and all the horizon quantities depend only on the same charge values.
The criticality condition (\ref{susyattractor}) gives $n_V$ complex independent conditions for $n_V$ scalar fields and hence fixes them all.

At the critical point the warp factor has a simple behaviour
\begin{equation}
	\label{warpbehaviour}
	(e^{-U})' = |Z|_*  \quad \Rightarrow \quad e^{-U} \to |Z|_* z,
\end{equation}
so that the near horizon metric approaches AdS$_2 \times S^2$.
Going back to the standard radial coordinate $r = -1/z$:
\begin{equation}
	ds^2 = - \frac{r^2}{|Z|_*^2} dt^2 + \frac{|Z|_*^2}{r^2} \left[dr^2 + r^2 d \Omega_{S^2}^2\right].
\end{equation}
The corresponding black hole entropy is given by the usual area formula, which in this case can be rewritten in terms of the central charge and in turn of the black hole potential at the horizon:
\begin{equation}
	\label{entrsusy}
	S_{BH} = \frac{A}{4}= \pi |Z|^2_* = \pi V_{BH}^*.
\end{equation}
Since the scalar fields at the horizon are all fixed in terms of the electric and magnetic charges of the solution, also the central charge
\begin{equation}
	|Z|_* = |Z|(p,q,z^i_*(p,q))
\end{equation}
depends only on the discrete charges and so does the entropy, according to (\ref{entrsusy}).

The geometric properties satisfied by the scalar manifold, namely the fact that it has a Special-K\"ahler nature, constrain the evaluation of the second derivatives of the central charge driving the supersymmetric flow so that 
\begin{equation}
	\partial_i \overline \partial_{\bar \jmath} |Z| = g_{i \bar \jmath} |Z| >0.
\end{equation}
This means that the critical points at which the flow stops are all minima of the central charge.
This also helps in understanding the attractor behaviour of the black hole horizon.
No matter what is the value at infinity of the scalar fields, they are driven by the flow equations towards the minimum of the central charge, which constitutes an attractor point for the differential equations determining the flow.
Eventually all the horizon quantities are determined by the value of the scalar fields at such minimum.
We can therefore think of our moduli space as a basin of attraction where the final attractor point, at the minimum of the basin, is specified only by the choice of the asymptotic charges.
\begin{wrapfigure}{l}{0.45\textwidth} 
  \vspace{-7mm}
\begin{center}
	\includegraphics[scale=1.1]{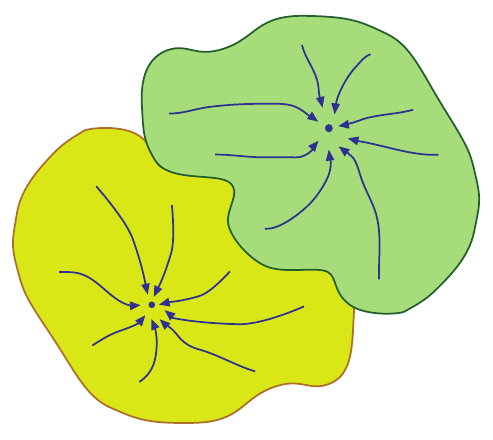} 
\end{center}
\caption{\small \emph{Representation of a moduli space with multiple basins of attraction.}} \label{basins} 
\end{wrapfigure}
We should note, however, that in some cases there can exist multiple basins of attraction, leading to a discrete number of possible values $z_*^i(p,q)$ for a given choice of charges.
In this case the attractor flow must be complemented by the ``area code'' corresponding to the basin of attraction to which the initial conditions belong \cite{Kallosh:1999mz,Wijnholt:1999vk,Kallosh:1999mb}.
For a SK manifold all these critical points will be minima of the central charge and there will be no other critical points, so that we are not in contradiction with the previous discussion.
If the reader wonders how such functions could be, an example with 2 minima and no other critical point in ${\mathbb R}^2$ is $f =  {e}^{-2x} - e^{-x -y^2}y^2$ \cite{Moore:1998pn}.

Before proceeding to the analysis of the non-supersymmetric case, let us analyze another interesting property of the flow equations (\ref{susyflow}): \emph{the supersymmetric c-theorem}.
It is a straightforward consequence of the previous analysis that the scale factor $\mu = e^{-U}$ is monotonically increasing along the flow
\begin{equation}
	\mu' = |Z| >0
\end{equation}
and therefore it can play the role of a c-function for the system (or rather its inverse), which has a minimum value at the Minkowski vacuum and blows up at the horizon (recall that $\mu_\infty = 1$ and $\mu_{hor} = +\infty$).
This implies that $\mu$ can replace the radial coordinate to describe the same flow.
By using the c-function as a parameter, the scalar field equations become a simple gradient flow equation:
\begin{equation}
	\mu \frac{d\ }{d\mu}z^i = g^{i\bar \jmath} \overline \partial_{\bar \jmath} \log |Z|.
\end{equation}

Here we focused on $N=2$ models, but supersymmetric solutions in models with $N>2$ follow essentially the same discussion where $|Z|$ is replaced by the largest of the absolute values of the eigenvalues of the central charge matrix.


\subsection{Non-BPS attractors} 
\label{sub:non_bps_attractors}

Although the solution of the Hamiltonian constraint given by (\ref{susyflow}) is straightforward and leads to supersymmetric attractors, the same theory (for different choice of charges) allows also for non-supersymmetric black holes.
These are described by critical points of the scalar potential $\partial_i V_{BH}^* = 0$ for which the central charge is not minimized $\partial_i |Z|_* \neq 0$.
The purpose of this section is to show that also in this case extremal black holes are described by first-order ordinary differential equations, driven by a function $W \neq |Z|$,  which we will call fake superpotential \cite{Ceresole:2007wx}.

There are two main motivations to believe that such a reduction may happen also for non-supersymmetric black holes.
The first one is that the attractor mechanism is at work also in this case.
Also for non-BPS extremal black holes the scalar potential drives the flow of the scalar fields in the moduli space towards the horizon value, which is once more specified by the minimum of the function.
This is also generically an attractor point, in the sense that critical points of $V_{BH}$ are generically minima, though for non-BPS black holes there exist flat directions.
An interesting remark in this case is that such flat directions, when they exist,  extend to the whole scalar potential and not just to the minimum.
Also, derivative corrections do not seem to destabilize the model \cite{Bellucci:2009nn}. 
The second one is that also non-BPS attractors have a c-theorem and therefore the flow has a clear and unique direction of motion in the moduli space.

The proof of the non-BPS c-theorem was given in \cite{Goldstein:2005rr}. 
We leave the reader to the original reference for the precise demonstration, while here we recall the general line of the argument.
The main assumption is that the matter involved in the theory giving the non-BPS extremal black holes satisfies the null energy condition.
This states that the stress energy tensor for such models should be positive definite when contracted with null vector fields:
\begin{equation}
	T_{\mu\nu} \zeta^\mu \zeta^\nu \geq 0, \qquad  \forall \ \zeta\ |\ \zeta^2 = 0.
\end{equation}
Once this assumption is made \cite{Goldstein:2005rr} proves that the area function $A$ decreases  monotonically along the solution towards the horizon and this means that one can identify such function with a c-function for the flow.
The demonstration uses the fact that the area function is a specific term appearing in the metric, once again also proportional to $e^{2U}$, whose derivatives can be identified with certain combinations of the Ricci tensor constructed from the same metric: $A' \sim -R_{rr} g^{rr} + R_{tt} g^{tt}$.
Now, using Einstein's equations and introducing a vector such that  $(\zeta^t)^2 = -g^{tt}$ and  $(\zeta^r)^2 = g^{rr}$, the same combination of the Ricci tensor describing the derivative of the area function can be rewritten as a combination of the stress energy tensor computed on the solution.
Finally, using the null energy condition this shows that the area is always decreasing along the flow
\begin{equation}
	A' \sim- R_{rr} g^{rr} +R_{tt} g^{tt} = -T_{\mu\nu} \zeta^\mu \zeta^\nu \leq 0.
\end{equation}

These facts and the analogy with a similar situation happening for non-BPS domain-wall solutions in the context of the gauge/gravity correspondence strongly suggests the existence of first-order equations also for non-BPS extremal black holes.
Within the AdS/CFT correspondence Renormalization Group flows of the dual field theory can be described in the gravitational setup as domain-walls interpolating between two different Anti-de Sitter vacua.
The field theory c-theorem guarantees a description of such domain-walls in terms of first order differential equations, also when there is no supersymmetry \cite{Skenderis:1999mm,Freedman:2003ax,Celi:2004st}.
Analogously, extremal black holes are solutions interpolating between Minkowski and AdS$_2 \times S^2$ vacua of the same model.
Using this analogy we can now make our argument solid.

Consider a simple model with a single scalar field $\phi$, subject to a scalar potential $V(\phi)$ admitting two different extrema.
An instantonic solution can be constructed by moving to euclidean signature and  assuming that the field depends only on one variable, so that $\phi'$ denotes its derivative with respect to such variable.
Different solutions are parameterized by different values of the energy and an extremal one is defined by solutions of the constrained equation of motion following from
\begin{equation}
	\label{520}
	{\cal L} = (\phi')^2 + V(\phi), 
	\qquad \qquad H =  (\phi')^2 - V(\phi) = 0.
\end{equation}
Although this system has a generic second order equation of motion, it is easy to see that a first-order equation is sufficient by using Bogomolnyi's trick of squaring the action:
\begin{equation}
	\label{action1st}
	S = \int dt\left(\phi' \pm \sqrt{V}\right)^2 \mp 2 \int dt \, \phi' \sqrt{V}.
\end{equation}
This action is equivalent to the one obtained by the previous Lagrangian, but now we clearly reduced the equation of motion for $\phi$ to a first order one 
\begin{equation}
  \phi' \pm \sqrt{V}=0.
\end{equation}
In fact, the second term in (\ref{action1st}) is always a total derivative and hence can be discarded, while the first term in brackets solves both the equation of motion and the Hamiltonian constraint.

The extension of this trick to the case where many scalars $\vec \phi$ are involved needs some care.
Extremal solutions of a system analogous to (\ref{520}) but with many scalars means solving the constrained equations of motion coming from
\begin{equation}
	\label{systvec}
	{\cal L} = |\vec\phi'|^2 + V(\vec \phi), \qquad H =  |\vec\phi'|^2 - V(\vec\phi) = 0,
\end{equation}
where the norm $|\vec \phi'|$ can be taken with respect to a positive definite metric also depending on the scalar fields.
The squaring of the action leads to
\begin{equation}
	S = \int dt\left|\vec\phi' \pm \vec n \,\sqrt{V}\right|^2 \mp 2 \int dt \,\vec n \cdot \vec \phi'\, \sqrt{V},
\end{equation}
where $\vec n$ is a unit-norm vector: $|\vec n|^2 = 1$.
In this example however the second term in the action is a boundary if and only if it is proportional to the field derivative of a new function ${\cal W}$:
\begin{equation}
	\vec n = \frac{\nabla_\phi {\cal W}}{\sqrt{V}}.
\end{equation}
We therefore conclude that the system of equations of motion and Hamiltonian constraint coming from (\ref{systvec}) can be described by first-order equations provided the scalar potential can be rewritten as the norm of the derivative of a scalar function:
\begin{equation}
	V(\phi) = |\nabla_\phi {\cal W}|^2.
\end{equation}
The Lagrangian and the Hamiltonian constraint of extremal black holes are a special instance of this general case, where the set of scalars comprises the moduli fields as well as the warp factor $\vec \phi =\{U,z^i\}$ and the metric defining the norm is factorized and equal to 1 in the $U$ direction and equal to $g_{i \bar \jmath}$ in the direction of the moduli fields.
Actually, given the special dependence on the warp factor, we can introduce a real valued \emph{fake superpotential} $W$ so that ${\cal W} = e^{U} W$ and the necessary constraint to rewrite the equations of motion in a first-order form reduces to
\begin{equation}
	e^{2U}V_{BH} = \partial_U(e^{U} W)^2 + 4\, \partial_i (e^U W) g^{i \bar \jmath} \overline\partial_{\bar \jmath} (e^U W),
\end{equation}
which implies that the black hole potential can be written as \cite{Ceresole:2007wx}:
\begin{equation}
	\label{VW}
	V_{BH} = W^2 + 4 \, \partial_i W g^{i \bar \jmath}\overline \partial_{\bar \jmath} W.
\end{equation}
Whenever this condition is satisfied, the black hole equations are reduced to first-order differential conditions \cite{Ceresole:2007wx}:
\begin{equation}
	\left\{\begin{array}{l}
	U' = - e^U W,\\[3mm]
	z^i{}' = - 2 \, e^U g^{i\bar \jmath} \, \overline \partial_{\bar \jmath}W.
	\end{array}\right.
\end{equation}
Once more the sign is fixed because of consistency of the behaviour of the warp factor from infinity to the horizon.
Obviously the BPS case is trivially recovered whenever $W = |Z|$, however, we will see that for a given black hole potential more solutions, with $W \neq |Z|$, can and will exist whenever there are extremal points of the black hole potential that are not extrema of the central charge.

The fake superpotential $W$ assumes a very important role for the description of the black hole solutions.
In fact, not only the flow equations determining the complete solution are driven by $W$, but also the mass and the entropy of the black hole are determined by the same function \cite{Ferrara:2008ap}.
In detail, the boundary term reduces to $e^U W$ and the value of the boundary term far away from the black hole determines the ADM mass: $e^U W \to W_{\infty} = M_{ADM}$.
Also in this case the flow equations will stop at the critical point of the function driving the scalars.
For non-BPS black holes this means that the horizon is reached whenever
\begin{equation}
	\partial_i W_* = 0.
\end{equation}
It is trivial to check that these are also critical points of the full black hole potential $V_{BH}$.
As discussed previously, in the non-BPS case there may be flat directions and actually this is reflected by $W$, which will not depend on the moduli related to such flat directions \cite{Ceresole:2009iy,Ceresole:2009vp,Andrianopoli:2010bj}.
From the flow equations we can also determine the behaviour of the warp factor close to the horizon, in full analogy with the supersymmetric case (\ref{warpbehaviour}), and this fixes the entropy to be
\begin{equation}
	S_{BH} = \frac{A}{4}= \pi W^2_* = \pi V_{BH}^*.
\end{equation}

We stress that the fake superpotential is an extremely powerful procedure that provides the full solution, including properties that depend on the behaviour of the scalar fields infinitely far away from the black hole, and not just the horizon properties as other procedures do.

Obviously the main problem connected with this technique is whether we can find any solution of the main constraint equation (\ref{VW}) other than the central charge of the system.
The answer is positive and two main techniques have been developed to provide such answer:
\begin{itemize}
	\item a constructive approach for coset manifolds based on duality invariants (see  \cite{Ceresole:2009iy,Ceresole:2009vp} and \cite{Andrianopoli:2007gt} also for $N>2$ theories);
	\item an  existence theorem in connection with the Hamilton--Jacobi equation \cite{Andrianopoli:2009je}.
\end{itemize}
The constructive approach is based on the simple observation that the warp factor $U$ is a duality invariant quantity (it is part of the metric and this is invariant under U-duality transformations).
Since the derivative of this function is related to $W$, also the fake superpotential must be invariant.
For a model based on a symmetric coset manifold describing the self-interactions of the vector multiplet scalar fields ${\cal M}_{sc} = G/H$, the duality group is $G \subset $Sp$(2n_V +2, {\mathbb R})$.
It is therefore a straightforward technical task to identify all possible invariants and find a working definition for $W$.
This can generically be done in few steps, by first identifying $W$ for a simple charge configuration, using symmetry properties to reconstruct the seed superpotential and then boost it by a duality transformation to generic charges.
One has to say that although the number of invariants is limited and one generically faces a well defined problem, the resulting fake superpotential may be extremely non-trivial, for instance non-polynomial in the invariants (see \cite{Bellucci:2008sv}) and hence the procedure can be technically challenging.
Also, this procedure does not apply to the cases where the scalar manifold is not a coset.

On the other hand, the Hamilton--Jacobi interpretation of our system of equations is very useful to derive a formal general solution and to prove an existence theorem, but it is also often unpractical in order to derive a closed expression for $W$.
The Hamilton--Jacobi equation is a first order nonlinear partial differential equation for a function ${\cal W}(\phi)$ called Hamilton's principal function such that
\begin{equation}
	{\cal H}\left(\phi^i,\frac{\partial {\cal W}}{\partial \phi^i}\right) = 0
\end{equation}
In our case the radial variable plays the role of a euclidean time, the fake superpotential plays the role of the principal Jacobi function while the set of all fields $\phi^i$ is assimilated to the coordinates of phase-space and the equation to be solved is defined by
\begin{equation}
	{\cal H} = V(\phi,Q) - \left|\frac{\partial {\cal W}}{\partial \phi}\right|^2.
\end{equation}
In this context ${\cal W}$ gets the interpretation of the generating function of canonical transformations of the classical Hamiltonian, so that $\pi^i = \frac{\partial {\cal W}}{\partial \phi^i}$ and $\pi^i = G_{ij} \phi^j{}' = \frac{\delta {\cal L}}{\delta \phi^i{}'}$ become the flow equations.
The existence of such a function is guaranteed by the Liouville integrability of the system \cite{Chemissany:2010zp,Fre:2011uy}.
From the general theory, ${\cal W}$ can also be formally constructed as the integral of the Lagrangian along the solution \cite{Andrianopoli:2009je}:
\begin{equation}
	{\cal W}(\phi) = {\cal W}_0 + \int_{\tau_0}^\tau {\cal L}(\phi,\phi') d \tau.
\end{equation}
Clearly this is unpractical in the generic situation where one does not know the solutions before having constructed ${\cal W}$.

An alternative technique has been proposed in \cite{Bossard:2009we}, where the superpotential is implicitly defined via the solution of an algebraic equation of degree 6, which follows from analyzing the geodesics in the time-like reduction of the black hole geometry.
This also gives a formal general definition of $W$, but the result of the solution of the equation of degree 6 is at least impractical.

\subsubsection{Examples} 
\label{sub:examples2}

The first example is given by the SU(1,1)/U(1) model with prepotential 
\begin{equation}
	F = -i X^0 X^1. 
	\label{prep} 
\end{equation}
The holomorphic sections and the K\"ahler potential were given in (\ref{sub:examples}).
For generic electric $q_\Lambda$ and magnetic $p^\Lambda$ charges, the central charge is then
\begin{equation}
	{Z} = \frac{q_0+i p^1 + (q_1 + i p^0) z}{\sqrt{2(z + \bar z)}}\, . 
	\label{Zcharge} 
\end{equation}
The black hole potential $V_{BH}$ is derived by inserting this expression in (\ref{VN2}): 
\begin{equation}
	V_{BH} = \frac{(p^1)^2-i {q_1} ({z}-{\bar z}) {p^1}+{q_0}^2+i {p^0} {q_0} ({z}-{\bar z})+\left((p^0)^2+(q_1)^2\right) {z} {\bar z}}{{z}+{\bar z}}. 
	\label{potenziale1modulo} 
\end{equation}
Black hole solutions are then found by looking for solutions interpolating between flat space at infinity and $AdS_2 \times S^2$ at the horizons defined by the critical points of $V_{BH}$.
Such critical points are found for 
\begin{equation}
	z^{\pm} = \frac{\pm (p^0p^1+q_0q_1)+i(p^0 q_0 - p^1 q_1)}{(p^0)^2 + (q_1)^2} , \label{zcrit} 
\end{equation}
and since consistency requires Re$z > 0$, they lie inside the moduli space only for $(p^0p^1+q_0q_1) >0$ when $z^+$ is chosen in (\ref{zcrit}), and for $(p^0p^1+q_0q_1)<0$ for $z^-$.
Different critical points have a different nature.
More precisely, $z^+$ (\ref{zcrit}) gives the supersymmetric vacuum, which satisfies $D_i Z = 0$, with ${ Z}\neq 0$, (hence $ \partial_i |{Z}| = 0$) and thus it is a fixed point of (\ref{Zcharge}), while the other critical point $z^-$ gives the non-BPS black hole, for which $D_i {Z} \neq 0$. 
The Hessian at these points is always positive as there are 2 identical positive eigenvalues 
\begin{equation}
	{\rm Eigen}\, \left\{{\rm Hess}(V_{BH})\right\} = \pm\frac{1}{p^0p^1+q_0q_1}\{((p^0)^2 + (q_1)^2)^2,((p^0)^2 + (q_1)^2)^2\} . 
	\label{eigenH} 
\end{equation}

A simple inspection of these formulae shows that the two type of black holes are related by a change of sign in the electric or magnetic charges. 
In fact, in this case, the fake superpotential is given by
\begin{equation}
	{\cal W}= \frac{\left|-q_0+i p^1 + (q_1 - i p^0) z\right|}{\sqrt{2(z + \bar z)}} ,\label{Zp} 
\end{equation} 
which indeed differs from (\ref{Zcharge}), but gives rise to the same potential $V_{BH}$. 
It is also quite simple to check that the critical point of this new ``fake superpotential'' is the non-BPS black hole, namely (\ref{zcrit}) with the minus sign. 
\begin{figure}
	[hbt] \centerline{ 
	\includegraphics[scale=1]{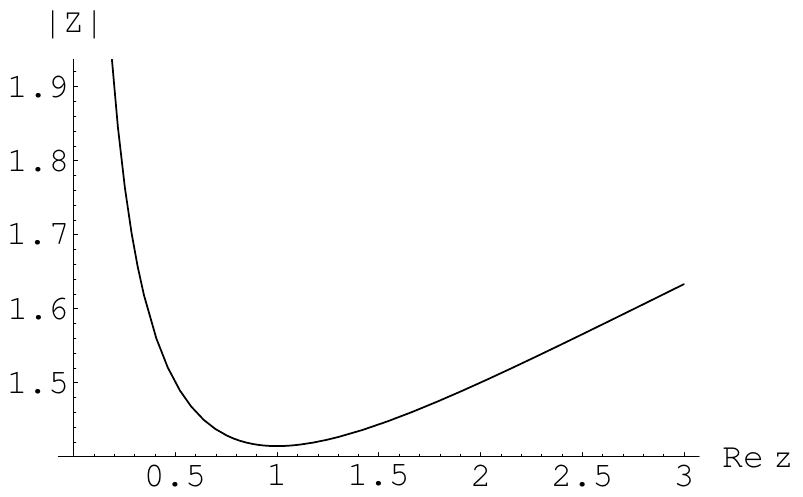} 
	\includegraphics[scale=1]{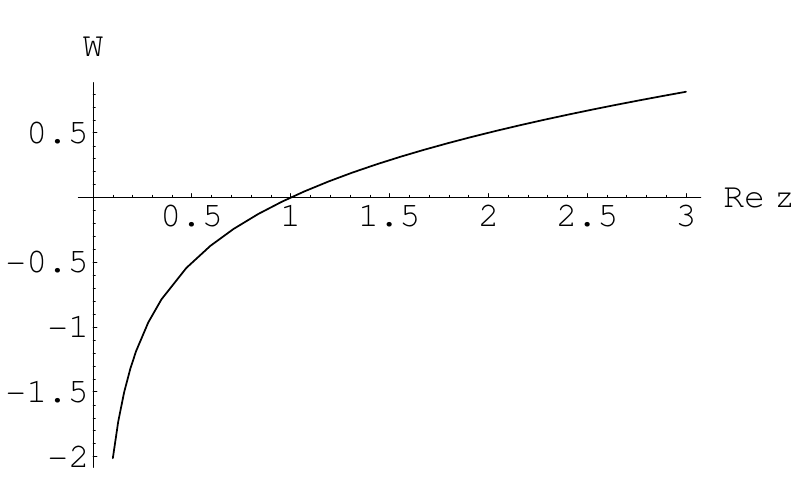} } \caption{Plots of the sections of $W$ and $|{\cal Z}|$ at Im $z=0$, for unit charges. 
	Where the central charge shows a minimum, the ``fake superpotential'' crosses zero. 
	Changing the signs of the $q_0$ and $p^0$ charges exchanges the two pictures.} \label{plots} 
\end{figure}

An extremely non-trivial example is the STU model, for which the complete superpotential expression can be found in \cite{Bellucci:2008sv}.
A simpler expression we can report here is the limiting case of the $t^3$ model, which essentially follows from the STU model by identifying $s=t=u$, for the case of only two non-trivial charges, $p^0$ and $q_1$.
The central charge is
\begin{equation}
	Z = \frac{z q_1 + p^0 z^3}{\sqrt{-i(z-\bar z)^3}},
\end{equation}
whose critical point is
\begin{equation}
	z^* = - i \sqrt{-\frac{q_1}{3 p^0}}.
\end{equation}
The corresponding fake superpotential is
\begin{equation}
	W = \frac{|z q_1 + p^0 z^2 \bar z|}{\sqrt{-i(z-\bar z)^3}},
\end{equation}
whose critical point is
\begin{equation}
	z^* = - i \sqrt{\frac{q_1}{3 p^0}}.
\end{equation}
This superpotential cannot be obtained from the central charge just by flipping charges.
Note also that when the supersymmetric critical point is well defined, the non-BPS critical point is not well defined and vice-versa.




\section{Duality} 
\label{sec:duality_orbits}

An important aspect of black hole solutions in supergravity is that the U-duality transformations map solutions of the various equations of motion and Bianchi identities to new solutions of the equations of motion and Bianchi identities of another system with different charges, preserving the metric.
This implies that the U-duality group $U$ can be used to generate solutions with arbitrary charges starting from more constrained configurations \cite{Hull:1994ys,Sen:1994eb,Cvetic:1995uj,Cvetic:1995bj,Cvetic:1996zq,Andrianopoli:1996ve,Ferrara:1997ci} by  
\begin{equation}
	\left(\begin{array}{c}
	p' \\ q'
	\end{array}\right) = S\left(\begin{array}{c}
	p \\ q
	\end{array}\right),
\end{equation}
where $S \in U \subset  {\rm Sp}(2 n_V + 2, {\mathbb R})$. 
We have also seen that such a technique has been used in order to provide a constructive mechanism to find the fake superpotential in the case of coset scalar manifolds.

An obvious interesting question is: what is the minimum number of charges that allows to generate arbitrary ones by a duality transformation?
The black hole solutions we considered so far are specified by the values of the charges $(p^\Lambda, q_{\Lambda})$ and by the asymptotic value of the scalar fields $z^i_{\omega = +\infty}$.
This gives a total of $2n_V +2 + 2n_V$ parameters specifying the complete solutions.
Note that while the first set is enough for defining the various quantities at the horizon, the second set is necessary for the full solution.
Consider now the case where the scalar manifold is given by a coset ${\cal M}_{sc} = G/H$.
The asymptotic values of the scalar fields $z^i_{\omega = +\infty}$ are parameters of the coset manifold and hence we can use the generators of $G/H$ to set them to whatever value we want.
The non-trivial parameters left are then $2 n_V + 2$, corresponding to the electric and magnetic charges of the theory.
We are also free to use duality transformations sitting in $H$, because they will not affect the value of the scalar fields\footnote{To be more precise: the scalar manifold can be parameterized by (right-invariant) coset representatives $L$, from which one can construct the scalar matrix $M = L^T L$ that enters in the scalar kinetic term. $M$ is invariant under $H$-transformations.}, but will rotate the charges.
Some $H$-generators may also have a trivial action on the charges, but all the others can be used to remove parameters of the solutions, which could be generated by the action of the duality group, instead.

An explicit example of this procedure can be outlined for the STU model.
A generic black hole in this model has 14 parameters: 8 charges and 6 real scalar fields.
The duality group is determined by the scalar manifold to be G = SU(1,1)$^3$, with H=U(1)$^3$.
As expected, the dimension of the coset coincides with the number of scalar fields and therefore the minimum number of parameters can be obtained by subtracting the dimension of $H$ to the total number of charges
\begin{equation} 
	\# {\rm charges} - \#({\rm H})  = 8- 3 = 5.
\end{equation}

A slightly more complicated example is given by N=8 supergravity.
This theory has 56 charges and 70 scalar fields for a total of 126 parameters for the generic black hole configuration.
The scalar manifold is E$_{7(7)}/$SU(8), and the dimensions of the two groups are $\#$[E$_{7(7)}$] = 133 and $\#$[SU(8)] = 63.
Surprisingly, if we straightforwardly apply the previous relation and subtract the number of $H$-generators to the number of allowed charges we would get a negative number.
However, the resolution of this little puzzle is rather simple if one analyzes more carefully the action of the $H$-group on the same charges.
In fact there is an $SU(2)^4 \subset H$ that leaves the charges invariant and therefore the total number of duality parameters we can use to reduce the number of independent charges is $\#(H) - \#[$SU(2)$^4]$ = 51.
This means that we are once more left with just 5 independent parameters to construct the seed solution that can be used to generate the most general one by duality.
This is not surprising after all, because the STU model analyzed previously is a special truncation of the $N=8$ model. 
It should also be noted that, although we fixed the scalars in order to determine the number of independent parameters, the seed solution could also be constructed by choosing any other 5 parameters among the charges and asymptotic scalars, with the obvious constraint that the invariant charge combination defining the entropy at the horizon should be non-vanishing (for large black holes).

Another important issue we will not analyze here in detail is the definition of the duality orbits \cite{Andrianopoli:1996ve,Ferrara:1997ci,Bellucci:2006xz}.
Since duality transformations can map black holes with different charges among themselves, it is useful to understand how many different orbits one has with respect to this action.
A full classification requires the construction of an appropriate number of independent invariant quantities.
For instance in N=8 supergravity different orbits are classified by the quartic invariant $I_4(p,q)$.
Whenever $I_4 >0$ one has supersymmetric black holes.
For $I_4 <0$ one has non-BPS configurations and, finally, when $I_4 = 0$, the horizon area vanishes and hence one has small black holes.
$N=2$ duality orbits were classified in \cite{Ceresole:2010nm}.


\section{Multicentre solutions} 
\label{sec:multicentre_solutions}

So far we concentrated our discussion on single centre black hole configurations.
However, a great deal of progress in our recent understanding of black hole physics within the context of String Theory came from multicentre solutions.
In this final section we will review such solutions with a special emphasis on the non-supersymmetric ones.

Once more, we would like to make contact with what has been discussed so far in the context of single centre configurations and therefore we will focus on ${ N} = 2$ theories in 4 dimensions.
In order to have an explicit relation with  String Theory it is actually very useful to see also how such configurations can be constructed in 10/11 dimensions within type II or M-theory models.
We chose to concentrate on M-theory compactifications on the product of a Calabi--Yau manifold and a circle: $CY \times S^1$.
This kind of compactification leads to an $N=2$ theory in 4 dimensions and it is also useful for to establish an explicit relation between quantities in 4 and 5 dimensions.
We will further specify the Calabi--Yau to be a simple orbifold 
\begin{equation}
	CY_6 = \frac{T^6}{{\mathbb Z}_2 \times {\mathbb Z}_2} \simeq (T^2)^3,
\end{equation} 
because this gives a reduced setup where the scalar manifold is described by the STU model.
In this scenario the 8 charges of the STU model correspond to different branes wrapped on the various cycles of the internal manifold.
Obviously, by replacing the internal manifold with a more general Calabi--Yau, we can get more general cubic Special--K\"ahler geometries, as discussed in section~\ref{sub:string_theory_origin}.
If we call $t$ and $\vec{x}$ the coordinates of 4-dimensional spacetime, $\psi$ the coordinate of the circle and $y^a$ the coordinates of the 6-torus, with the two ${\mathbb Z}_2$ orbifold actions inverting the sign of the first 4 and last 4 coordinates respectively, the charge configuration can be summarized by the following table:
\begin{equation}
		\renewcommand{\arraystretch}{1.4}
	\begin{array}{|c|ccccccccccc|c|c|}\hline\hline
	 {\rm M-theory} & t & x_1 & x_2 & x_3 & \psi & y_1 & y_2 & y_3 & y_4 & y_5 & y_6 & IIA & charge \\\hline\hline
	KK6 & - & &&& -& -& -& -&-&-&-& D6 &p^0\\\hline
	M5 & - &&&  & -& -& -& -& -&&& D4 &p^1 \\\hline
	M5' & - &&& & -& -& -& & &-&-& D4' &p^2 \\\hline
	M5'' & - &&&  &- & & & -& -&-&-& D4'' &p^3 \\\hline
	M2 &- &&&&&-&-&&&&&D2&q_1\\\hline
	M2' &- &&&&&&&-&-&&&D2'&q_2\\\hline
	M2'' &- &&&&&&&&&-&-&D2''&q_3\\\hline
	KK0 &- &&&&-&&&&&&&D0&q_0\\\hline\hline
	\end{array}
\end{equation}
where the line denotes the extension of the brane along that direction.
From the M-theory point of view we have 6 real charges ($p^i$ and $q_i$) and two geometric charges ($p^0$ and $q_0$).
The first one arises from Kaluza--Klein monopoles in M-theory, and can be seen as an additional brane charge in 10 dimensions (a D6-brane charge), while $q_0$ arises as the charge related to the 4-dimensional vector arising from the 5-dimensional component of the metric describing the fibration of $\psi$ on $\vec x$ and corresponds to a Kaluza--Klein particle (with a nontrivial momentum along $\psi$).

It is actually better to describe the reduction process to 4 dimensions in 2 steps.
First of all we consider the reduction to 5 dimensions along the $(T^2)^3$ and then discuss further reductions to 4 dimensions.
From the M-theory point of view a stationary metric ansatz that takes into account the backreaction of the above configuration of branes on the geometry is
\begin{equation}
	ds_{11}^2 = - Z^{-2} \left(dt+\omega \right)^2 + Z \, ds_4^2(x) + \sum_I \frac{Z}{Z_I}\, ds_{T^2}^2.
\end{equation}
Here $ds_4^2$ is a Ricci-flat 4-dimensional euclidean space, which can be chosen to be ${\mathbb R}^4$, the Gibbons--Hawking space, the Euclidean Schwarzschild metric or others according to the type of solution we want to describe.
The rest of the ansatz is chosen so that the total volume of the internal manifold remains fixed, which implies that no hypermultiplets will be turned on in the lower-dimensional effective theory.
The warp factors also depend only on the coordinate of the 4-dimensional euclidean space and 
\begin{equation}
	Z = \left(Z_1 Z_2 Z_3\right)^{1/3}.
\end{equation}
Finally, the M-theory ansatz is completed by the 3-form potential, which we take as
\begin{equation}
	C_3 = \sum_I \left(\frac{-dt + \omega}{Z_I}+a_I\right)\wedge dT_I,
\end{equation}
where $dT_I$ is the volume element of the $I$-th torus.
This gives the so-called \emph{floating brane ansatz}.
The name follows from the fact that any probe M2 brane wrapping one of the 2-tori feels no force thanks to the cancellations between the contributions coming from the Dirac--Born--Infeld action and the Wess--Zumino terms due to the same dependence of the metric and $C_3$ on the warp factors $Z_I$.

By using this ansatz, the Bianchi identities and equations of motion of the 11-dimensional theory reduce to  almost linear equations depending only on the coordinates of $ds_4^2(x)$ \cite{Bena:2004de}:
\begin{eqnarray}
	\label{eqs1}
	da_I &=&  \star_4 da_I, \\[2mm]
	d \star_4 dZ_I &=& \frac{|\epsilon_{IJK}|}{2} da_J \wedge \star_4 da_k, \\[2mm]
	\label{eqs3}
	d \omega + \star_4 d \omega &=& Z_I d a_I.
\end{eqnarray}
\begin{wrapfigure}{r}{0.55\textwidth} 
\vspace{-9mm}
\begin{center}
	\includegraphics[scale=.15]{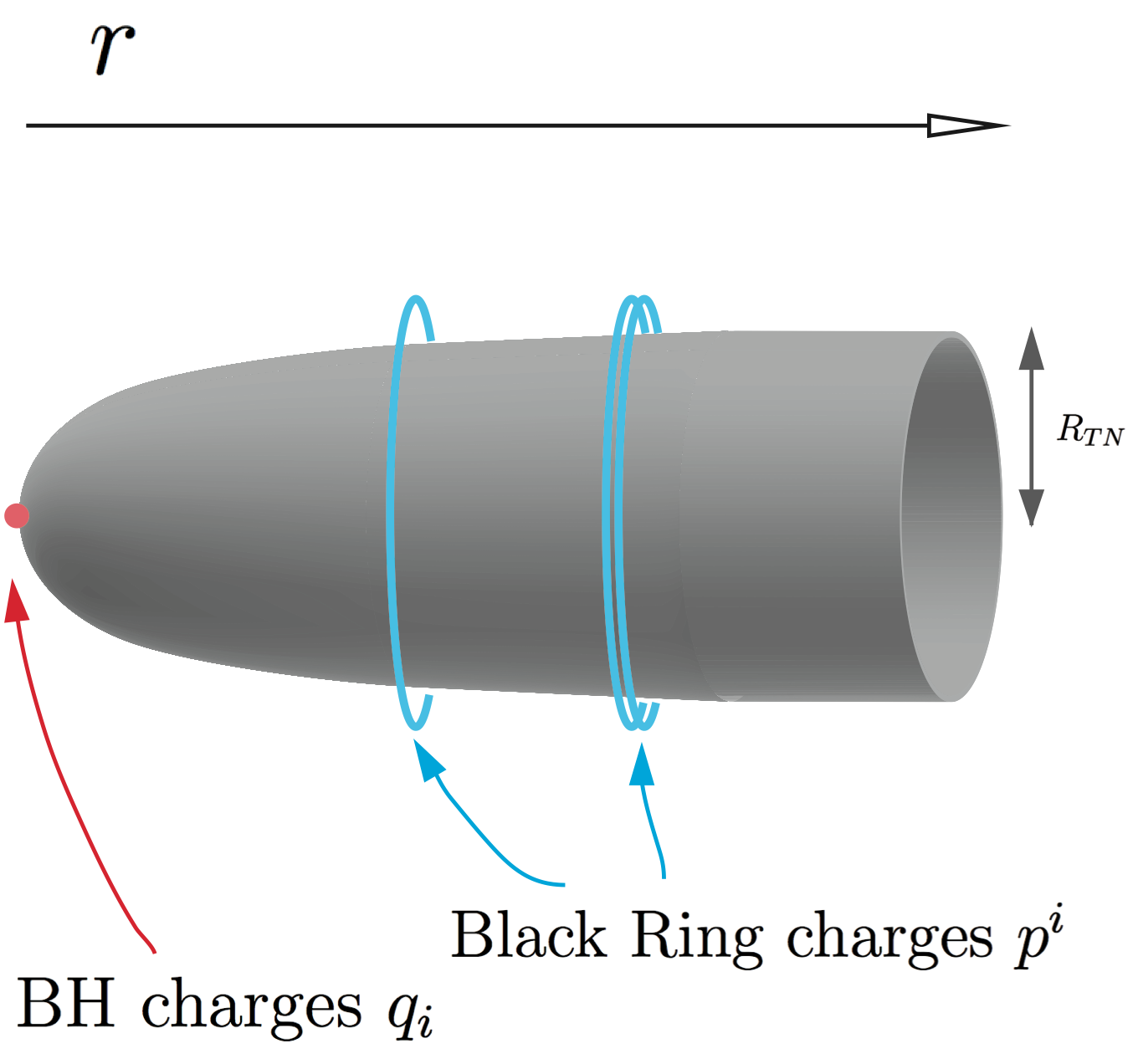} 
\end{center}
\vspace{-7mm}
\caption{\small \emph{BH's in a GH geometry.}} \label{taubnut} 
\end{wrapfigure}
As discussed above, in the end we would like to describe black hole geometries in 4 dimensions.
This means that we need to choose the metric of $ds_4^2$ as a Ricci-flat circle fibration on a 3-dimensional base, so that we can reduce the model along the circle direction and go back to the STU model.
The first choice we will consider is that of a Gibbons--Hawking space.
This is a 4-dimensional euclidean space endowed with a metric
\begin{equation}
	ds^2_4 = \frac{1}{V}\left(d \psi +\vec A\right)^2 + V d\vec x_3^2,
\end{equation}
where 
\begin{equation}
	\label{signTN}
	\star d\vec A = \pm dV.
\end{equation}
We assume that $\psi$ is a U(1) isometry and hence $V$ depends only on the coordinates of the 3-dimensional flat base $d\vec x_3^2 = dx_1^2 + dx_2^2+dx_3^2$.
By consistency, $V$ is a harmonic function which generates the NUT charge corresponding to the presence of a D6 brane in 10 dimensions: 
\begin{equation}
	V = h^0 + \frac{p^0}{r}.
\end{equation}
The overall geometry of this space is that of a cigar (Look at Picture~\ref{taubnut}).
At the tip ($r \to 0$) one has the NUT charge $p^0$ and spacetime looks 5-dimensional.
At large values of $r$, the Gibbons--Hawking space becomes the direct product of ${\mathbb R}^3$ with a circle of radius 
\begin{equation}
	R_{TN} = \frac1{\sqrt{h^0}}.
\end{equation}

The choice of sign in (\ref{signTN}) is equivalent to a choice of orientation on the space and has a dramatic effect on the reduction of the equations of motion (\ref{eqs1})--(\ref{eqs3}).
Depending on the sign one gets two different sets of equations which correspond to supersymmetric and non-supersymmetric configurations \cite{Goldstein:2008fq,Bena:2009ev}.
In detail, the plus sign gives BPS configurations, whereas the minus sign breaks supersymmetry.
The supersymmetry breaking is obviously mild and is essentially just a reaction to the change in the global conditions (the orientation of the space).
For this reason these conditions have been named \emph{almost BPS} \cite{Goldstein:2008fq}.

The knowledge of the map between the quantities entering in the definition of the 11-dimensional ansatz and the fields defining the 4-dimensional STU supergravity model allows the use of the 4-dimensional duality group to generate new solutions starting from known ones.
The full map is given in \cite{Dall'Agata:2010dy}.
However, we show here how the reduction along the $\psi$ circle allows the definition of new quantities in terms of which one defines the physical scalars in 4 dimensions.
Now quantities with a vector have legs only along $d\vec x_3^2$.
The 1-forms appearing in the metric and M-theory potential now split as
\begin{eqnarray}
	a_I &=& C^I( d \psi - \vec A^0) + \vec A^I, \\[2mm]
	\omega &=& \mu (d \psi - \vec A^0) + \vec \omega,
\end{eqnarray}
so that the scalar fields of the STU model are parameterized by
\begin{equation}
	z^I = C^I - i \frac{X^I}{\Delta^2},
\end{equation}
where $X^I = Z/Z_I$ is the volumes of the I-th torus and $\Delta$ is a factor that depends on the fibration. 
For the Gibbons--Hawking reduction \cite{Dall'Agata:2010dy}
\begin{equation}
	\Delta^4 = \frac{Z^{2/3}V}{V Z^3 - V^2 \mu^2}.
\end{equation}
As explained in section~\ref{sub:examples}, each coordinate spans a SU(1,1)/U(1) factor and the U-duality group is [SU(1,1)]$^3$.
The action of the duality group on the scalar fields can be represented as a fractional linear transformation
\begin{equation}
	z^i \to \frac{a_i z^i + b_i}{c_i z^i + d_i},
\end{equation}
where the parameters are part of an SU(1,1) valued matrix:
\begin{equation}
	M_i = \left(\begin{array}{cc}
	a_i & b_i \\
	c_i & d_i
	\end{array}\right) \in SU(1,1)_i.
\end{equation}
We stress that each of these transformations acts also on the charges, but leaves the 4-dimensional metric invariant.
On the other hand, such transformations do not leave the 5-dimensional or 11-dimensional metric invariant and this may have profound consequences on the form of the solution as seen from M-theory.
Although some of the 4-dimensional duality transformations have an obvious 11-dimensional interpretations, other become rather non-trivial and involved in the uplifting process.
For instance, the combination of 2 T-duality transformations on the I-th torus gives a matrix transformation of the form
\begin{equation}
	M_I = \left(\begin{array}{cc}
	0 & -1 \\
	1 & 0
	\end{array}\right)
\end{equation}
on the corresponding scalar.
It is interesting to note that from the 4-dimensional point of view this is an S-duality transformation $z \to -1/z$.
Gauge shift symmetries of $C_3$ along the tori also have a straightforward representation:
\begin{equation}
	M_I = \left(\begin{array}{cc}
	1 & \lambda_I \\
	0 & 1
	\end{array}\right).
\end{equation}
On the other hand, the action of 
\begin{equation}
	M_I = \left(\begin{array}{cc}
	1 & 0 \\
	\lambda_i & 1
	\end{array}\right)
\end{equation}
has the interpretation of a spectral flow transformation and only the rewriting in terms of 4-dimensional duality transformations allowed us to identify such a transformation with the combination of 6 T-dualities, a gauge transformation and 6 inverse T-dualities.
The same duality transformations act also on the charge vectors as described in \cite{Behrndt:1996hu,Dall'Agata:2010dy}.

\subsection{BPS case} 
\label{sub:bps_case}

The 4-dimensional black hole solution is generically described by the warp factor $U$, by the scalar fields $z^i$ and by the time components of the electric and magnetic vector fields $\{A^\Lambda, A_{\Lambda}\}$, whose duals we replaced with the charges in previous examples.
In the multi-centre case, the solutions will not be static anymore and therefore one also needs the 1-form $\vec \omega$.
Using the setup considered in this section, the electric vector fields follow from the reduction as $A^\Lambda \sim \left\{\vec A^0, \vec A^I\right\}$.

In the case of the choice of positive orientation on the Gibbons--Hawking metric, the solution is supersymmetric.
The full solution was first found in \cite{Denef:2000nb} and it can be completely expressed in terms of 8 harmonic functions
\begin{eqnarray}
	\star_3 d\vec A^\Lambda &=& d H^\Lambda, \\[2mm]
	\star_3 d\vec A_\Lambda &=& d H_\Lambda,
\end{eqnarray}
with
\begin{equation}
	H = h + \sum_i \frac{Q_i}{|\vec x - \vec x_i|},
\end{equation}
where $Q_i$ represent the appropriate electric $q_{\Lambda\,i}$ or magnetic charge $p^\Lambda_i$.
Also the non-static part of the metric is given in terms of the same harmonic functions
\begin{equation}
	\star d\vec \omega = \langle dH, H\rangle
\end{equation}
and its existence is related to the intrinsic angular momentum due to the electric-magnetic field generated by the static charges.

Consistency also implies that the positions of the charges is constrained by the equations
\begin{equation}
	\sum_j \frac{\langle Q_i,Q_j\rangle}{|\vec x_i - \vec x_j|} = 2 {\rm Im}\, (e^{-i \alpha} Z(q_i))_\infty,
\end{equation}
whose zeros give the positions in moduli space related to marginal stability of the solution (some of the distances blow up and therefore components are not bound anymore).

The application of duality transformations to this solution simply rotates the harmonic functions among themselves \cite{Dall'Agata:2010dy}.
This means that for a supersymmetric solution the 11-dimensional metric and 3-form potential are always described by an ansatz like the one we presented above, though the details in terms of the charges depend on the choices of the harmonic functions.


\subsection{non-BPS case} 
\label{sub:non_bps_case}

The non-supersymmetric case is more interesting (at least for what concerns the construction of solutions).
In this case not all possible solutions will fall in the ansatz above.
Actually many interesting new solutions that are candidate microstate geometries for black holes can be obtained by different bubbling equations.
These depend on the form of our choice of $ds_4^2$ and consequently on the choice of the 3-form potential and warp factors.
Starting from the Gibbons--Hawking space (Almost BPS solutions), the bubbling equations become
\begin{eqnarray}
	d\vec A^I &=& C^I dV - V dC^I, \\[2mm]
	d \star_3 dZ_I &=& \frac{|\epsilon_{IJK}|}{2}\, V d \star_3 d(C^J C^K), \\[2mm]
	\star_3 d \omega &=& d (\mu V) - V Z_I d C^I.
\end{eqnarray}
In opposition to what happened in the case of supersymmetric solutions one cannot generally solve these equations only in terms of harmonic functions.
If one insists in doing so, then only mutually local solutions exist and the positions of the various centers are not constrained \cite{Gaiotto:2007ag,Gimon:2009gk}.
However one can find more general solutions with mutually non-local charges and constrained positions\footnote{The apparent discrepancy between the results in \cite{Gaiotto:2007ag,Gimon:2009gk} on the one hand and \cite{Bena:2009ev} on the other is due to the different implementation of the regularity requirements.
In order to have regular horizons, the warp factor should not grow too fast when approaching the horizon.
If this behaviour is constrained by considering the asymptotic charges, the only possible regular solutions are all marginally stable and the solution is given entirely in terms of harmonic functions. 
On the other hand this global requirement is not necessary and application of this condition centre by centre allows for regular bound states as in \cite{Bena:2009ev}}.

Before moving to the multi-centre case, we can see that from the above ansatz one can easily recover the single centre seed solution with D2 and D6 charges and a non-trivial axion \cite{Lopes Cardoso:2007ky,Gimon:2007mh,Bena:2009ev} (this is a total of 4 charges and one non-trivial asymptotic value for the scalar fields).
This corresponds to setting $C^I =0$.
The bubbling equations imply that now $V$ and $Z_I$  are harmonic functions and $\mu = \frac{b}{V}$, where $b$ is the asymptotic value of the axions.
The warp factor of the black hole metric follows as
\begin{equation}
	e^{-4U} = 4 V Z_1 Z_2 Z_3 - b^2.
\end{equation}

For a multicentre solution we can start by requiring $C^I$ to be harmonic functions.
This, in turn, implies that the $Z_I$ are not harmonic anymore.
The full solution is rather complicated and can be found in \cite{Bena:2009ev}.
We stress that regularity of the solution in this case implies the existence of non-trivial constraints already for 2 centres.
This is actually an equation for the distance between the centres that is given in terms of charges and asymptotic values of the scalar fields.

The existence of a constraint on the distances implies that such solutions describe bound states of black holes.
It is obviously of much interest to understand the general conditions under which such bound states exist depending on the charges and the point in moduli space, like in the BPS case, where such an analysis culminated in the wall crossing formulae (See for instance the lectures at \cite{Denef:2007vg,Moorelectures}).
A first attempt at an analysis in this direction is presented in \cite{Ferrara:2010cw,Ferrara:2010ru}, but only in special models and for 2-centre solutions.

Once more the solution obtained in \cite{Bena:2009ev} can be used as a seed to obtain more general ones by using duality transformations.
However, in this instance, the action of the duality group changes the 5-dimensional metric \cite{Dall'Agata:2010dy}.
We show here a simple example.
Choose $C^1 = C^2 = 0$ and $C^3 \neq 0$, which means only one non-trivial D4-charge.
By inspection of the bubbling equations we now see that the 11-dimensional warp factors $Z_I$ are harmonic and only $\mu$ cannot be given in a closed form in terms of harmonic functions, because of its equation, which is
\begin{equation}
	d \star d (V \mu) = d (V Z_3)\wedge \star \,d C^3.
\end{equation}
Performing 6 T-dualities, following the approach described in the previous section, the new 5-dimensional metric contains a new spatial part, which describes now an Israel--Wilson space, rather than a Gibbons--Hawking.
The dual metric is
\begin{equation}
	ds_4^2 = (V_1 V_2)^{-1}(d \psi - \vec A)^2 + V_1 V_2 \,d \vec x_3^2,
\end{equation}
where $V_1 = C^3$, $V_2 = Z_3$ and $\star d\vec A = V_2 \,dV_1 - V_1\, d V_2$.
It is obvious that this will never be of the Gibbons--Hawking form.

The fact that the floating brane ansatz encompasses all BPS solutions, whereas it does not contain all non-BPS solutions, implies that the non-supersymmetric configurations are far more richer than the supersymmetric ones.
Unfortunately, this also means that the attempts at constructing a general solution, valid in any frame, as was done for the BPS multicentre black holes in \cite{Denef:2000nb}, have to face harder challenges.
Some progress in a 4-dimensional setup was made in \cite{Galli:2010mg}.
Moreover, having more centres means that duality orbits are classified by a larger number of invariants \cite{Ferrara:2010cw} and the non-BPS ones of \cite{Bena:2009ev} are the seed solutions only for one of the orbits in \cite{Ferrara:2010cw}.

Similar techniques can be used to generate more general new classes of non-BPS solutions, however it is very difficult to find a sufficiently general explicit solution that fulfills the constraint on the positions.
The best obtained so far was a line of rotating black holes \cite{Bena:2009en}.

Before wrapping up, let us also mention that such bubbling equations are also very useful to obtain other general solutions whenever $ds_4^2$ has a more general form.
In particular one can obtain new 5-dimensional smooth solutions with the same charges of 4-dimensional black holes and hence interpret these states as candidate microstate geometries \cite{Bena:2009fi,Bena:2009qv,Bena:2010gg}.
 
Also, the reduction to 4-dimensions of these stationary solutions allows also for rotating configurations in 4-dimensions as well.
In fact, by this mechanism we could find the most general seed solution for slowly rotating black holes ($a \to 0, m\to 0$, $J = a/m$ fixed) \cite{Bena:2009ev}.
The overrotating counterpart ($a=m \neq 0$) is still missing because of the different structure of the ansatz.

As a final note we would like to note that the idea of reducing extremal black hole equations to a first-order formalism \cite{Ceresole:2007wx} has revised the search for exact solutions also in other instances where some of the conditions assumed in \cite{Ceresole:2007wx} are relaxed.
Recently we have seen this formalism applied to supersymmetric black holes in U(1) gauged supergravity \cite{Dall'Agata:2010gj}  (solutions in this context were also obtained in \cite{Bellucci:2008cb,Cacciatori:2009iz,Hristov:2010ri}), to rotating extremal solutions \cite{Galli:2010mg} and even to classes of non-extremal black holes \cite{Andrianopoli:2009je,Perz:2008kh,Galli:2011fq}.




\bigskip
\section*{Acknowledgments}

\noindent The original parts of the contents of these lectures come from collaborations with I.~Bena, A.~Ceresole, S.~Ferrara, S.~Giusto, G.~Lopes Cardoso, J.~Perz, C.~Ruef, C.~Toldo, A.~Yeranyan and N.~Warner, which are gratefully acknowledged.
I would also like to thank the organizers of the ``27th Nordic Spring String Meeting'', ``SAM 2009'' and ``BOSS 2011'' schools for the kind hospitality and the nice and stimulating environment.
This work is supported in part by the ERC Advanced Grant no. 226455, \textit{``Supersymmetry, Quantum Gravity and Gauge Fields''} (\textit{SUPERFIELDS}), by the Fondazione Cariparo Excellence Grant {\em String-derived supergravities with branes and fluxes and their phenomenological implications}, by the European Programme UNILHC (contract PITN-GA-2009-237920) and by the Padova University Project CPDA105015/10.


\end{document}